\documentclass[
  reprint,
  superscriptaddress,
  amsmath,
  amssymb,
  aps,
  prb,
  longbibliography,
  floatfix
]{revtex4-2}

\usepackage{graphicx}
\usepackage{bm}
\usepackage{braket}
\usepackage{booktabs}
\usepackage{xcolor}
\usepackage[colorlinks=true,allcolors=blue]{hyperref}

\newcommand{\mub}{\mu_{\mathrm B}}
\newcommand{\ueV}{\mu\mathrm{eV}}
\newcommand{\dd}{\mathrm d}

\newcommand{\Tr}{\operatorname{Tr}}

\begin{document}

\title{Symmetry-Selective Strain Control of Anisotropic Magnetic Response in a Silicon FinFET Double Quantum Dot}

\author{Yuze Lu}
\affiliation{School of Integrated Circuits, Peking University, Beijing 100871, China}
\author{Xiaoyan Liu}
\affiliation{School of Integrated Circuits, Peking University, Beijing 100871, China}
\affiliation{Beijing Advanced Innovation Center for Integrated Circuits, Beijing 100871, China}
\author{Fei Liu}
\email{feiliu@pku.edu.cn}
\affiliation{School of Integrated Circuits, Peking University, Beijing 100871, China}
\affiliation{Beijing Advanced Innovation Center for Integrated Circuits, Beijing 100871, China}

\begin{abstract}
Strain naturally develops in three-dimensional quantum-dot structures such as silicon FinFETs during fabrication and cooling.
Such strain becomes especially important in a double quantum dot, because the two dots can experience different local strain and therefore acquire different magnetic responses.
To understand how this dot-to-dot strain difference affects coupled hole spins, we theoretically study the local \(g\) tensors of a silicon FinFET double quantum dot by combining a three-dimensional Poisson--Schr\"odinger calculation based on a six-band \(k\!\cdot\!p\) model with configuration interaction.
We find that the effect of strain depends on both its tensor component and its spatial symmetry.
For the diagonal components \(\epsilon_{yy}\) and \(\epsilon_{zz}\), strain mainly changes the principal \(g\) values, with only a small opening of the maximum-response axes.
In contrast, the shear component \(\epsilon_{yz}\) can also change the orientation of the local magnetic response.
When the strain profile preserves the transverse mirror symmetry, the shear-induced rotation is strongly suppressed.
Breaking this local constraint permits a pronounced off-diagonal response and rotates the principal magnetic axes.
The same component- and symmetry-selected trends appear in a Zeeman-only calculation, showing that the valence-band Zeeman coupling is sufficient to generate them, while the full Hamiltonian determines their quantitative expression.
Together, these results show how the tensor component and spatial symmetry of strain can be used to control both the magnitude and orientation of the magnetic response in coupled hole-spin qubits.
\end{abstract}

\maketitle
\flushbottom

\section{Introduction}
\label{sec:introduction}

Semiconductor quantum-dot spins combine well-isolated quantum states with local electrical control, making them a leading route toward scalable quantum processors~\cite{LossDiVincenzo1998,Zwanenburg2013,Burkard2023}.
Silicon is especially attractive because isotopic enrichment suppresses nuclear-spin noise, while its compatibility with complementary metal--oxide--semiconductor (CMOS) processing offers a route to dense and reproducible devices~\cite{Zwanenburg2013,Zwerver2022,Neyens2024}.
Silicon electron quantum dots have consequently progressed from high-fidelity single-spin control to two-qubit gates and multiqubit processors~\cite{Yoneda2018,Xue2022,Noiri2022,Philips2022}.

Hole spins provide a complementary route in which the intrinsic valence-band spin--orbit interaction enables electrical spin manipulation without an added magnetic-field gradient~\cite{Maurand2016,Piot2022,Camenzind2022}.
The same interaction makes their Zeeman response strongly anisotropic and electrically tunable~\cite{Ares2013,Crippa2018,Venitucci2018,Liles2021}.
This anisotropic low-field response is naturally described by a \(g\) tensor: its principal values give the response strengths, and its principal axes give the corresponding magnetic-field directions.
Because the underlying valence-band admixture depends on confinement, the \(g\) tensor is sensitive to device geometry, crystal orientation, and electrostatic tuning~\cite{Bosco2021,Martinez2022,Malkoc2022,Wang2024}.
FinFETs are particularly relevant in this setting: their wrap-around gates provide strong three-dimensional confinement and compact transistor-compatible control, and silicon hole-spin operation has already been demonstrated in this geometry~\cite{Maurand2016,Camenzind2022,Fuhrer2022}.

The same three-dimensional, multi-material structure also makes strain unavoidable.
Small inhomogeneous strains in silicon hole devices have been attributed to device processing and cooldown~\cite{Piot2022}, while thermoelastic simulations show that differential thermal contraction produces spatially varying strain across a FinFET~\cite{Bouquet2025}.
By changing the heavy-hole, light-hole, and split-off-band admixtures that underlie the anisotropic magnetic response, strain can reshape the local \(g\) tensor~\cite{Liles2021,Wang2024}.
Spatially varying strain can additionally generate spin--orbit interactions of its own~\cite{AbadilloUriel2023}.

In a double quantum dot (DQD), the two localized holes occupy different positions in this strain field and may therefore acquire distinct local tensors \(\bm G_L\) and \(\bm G_R\).
Silicon hole DQDs have been realized in CMOS-compatible devices~\cite{Ezzouch2021,Jin2023}, and site-dependent hole \(g\) factors have been resolved directly by magnetospectroscopy~\cite{Russell2023}.
For exchange-coupled spins, a difference between the local magnetic responses changes singlet--triplet mixing and makes the two-hole spectrum depend on the magnetic-field direction~\cite{Hwang2017,Hetenyi2020,Sen2023,Geyer2024,Liles2024}.

The question addressed here is how the tensor component of strain and the transverse spatial symmetry of its profile within a silicon FinFET DQD select the form of this local \(g\)-tensor difference.
The two tensors may differ in their principal values, in the orientation of their principal axes, or in both, with distinct consequences for the correlated two-hole spectrum.

We answer this question with a self-consistent three-dimensional Poisson--Schr\"odinger calculation based on a six-band \(k\!\cdot\!p\) model, followed by configuration interaction (CI) and a two-site Fermi--Hubbard reduction.
We find that the diagonal strain components \(\epsilon_{yy}\) and \(\epsilon_{zz}\) predominantly change the principal \(g\) values, with little opening of the maximum-response axes.
For the shear component \(\epsilon_{yz}\), a mirror-compatible \(z\)-odd profile suppresses the local \(xz/yz\) response at each dot, whereas a mirror-breaking \(z\)-even profile generates a pronounced off-diagonal response and rotates the principal axes.
The resulting local-tensor changes are directly reflected in the correlated two-hole spectra.
Finally, a calculation containing only the valence-band Zeeman magnetic coupling reproduces the leading strain and symmetry trends, identifying a microscopic route by which they arise.

\section{Device and numerical framework}
\label{sec:methods}

\subsection{Device geometry}
\label{sec:device_method}

\begin{figure}[!t]
  \centering
  \includegraphics[width=\columnwidth]{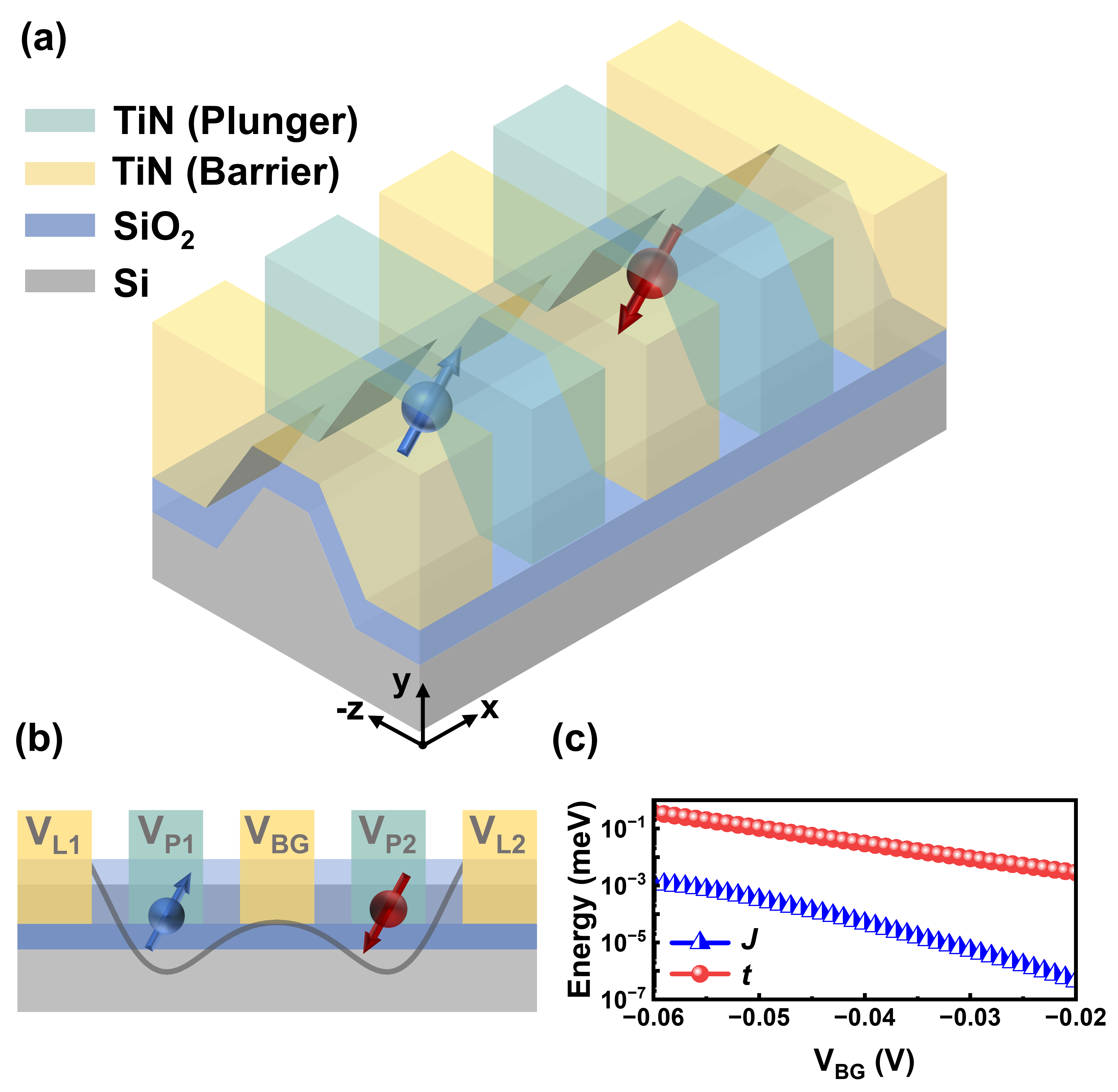}
  \caption{\label{fig:device}
  Silicon FinFET double quantum dot and electrical tuning.
  (a) Three-dimensional device schematic with the two localized hole states.
  (b) Longitudinal confinement profile and gate assignment.
  (c) Calculated two-hole singlet--triplet splitting \(J\) (blue triangles) and single-hole tunnel coupling \(t_c\) (red circles, labeled \(t\) in the panel) as functions of barrier-gate voltage at zero plunger detuning for the unstrained reference \(\bm\epsilon(\bm r)=0\).
  All plotted points are numerical results.}
\end{figure}

The simulated structure is the silicon FinFET double quantum dot sketched in Fig.~\ref{fig:device}.
Two plunger gates, \(P1\) and \(P2\), define the confinement minima.
The central barrier gate \(BG\) controls interdot tunneling, and the outer gates \(L1\) and \(L2\) represent the reservoir-gate sections.
Gate voltages enter as Dirichlet boundary conditions at the \(\mathrm{SiO_2}\) gate-oxide boundary.
The silicon fin has a trapezoidal cross section with bottom width \(W_{\mathrm{bottom}}=34~\mathrm{nm}\), top width \(W_{\mathrm{top}}=4~\mathrm{nm}\), and height \(H=22~\mathrm{nm}\).
Along the fin, \(L_{\mathrm{LG}}=30~\mathrm{nm}\), \(L_{\mathrm{PG}}=15~\mathrm{nm}\), \(L_{\mathrm{BG}}=30~\mathrm{nm}\), and \(L_{\mathrm{gap}}=5~\mathrm{nm}\).
The oxide thickness is \(10~\mathrm{nm}\).
The fin axis is \(x\parallel[110]\), the growth direction is \(y\parallel[001]\), and the transverse direction is \(z\parallel[1\bar 10]\).

\subsection{Single-hole calculation}
\label{sec:single_hole}

The single-hole states are obtained from a three-dimensional finite-element Poisson--Schr\"odinger calculation based on the standard six-band Luttinger--Kohn \(k\!\cdot\!p\) Hamiltonian, with the electrostatic confinement supplied by the Poisson solution.
The full magnetic Hamiltonian is written as
\begin{equation}
H_{\mathrm{full}}
=H_{\mathrm{LK}}(\bm{\pi})
+H_{\mathrm{PB}}(\bm{\epsilon})
+V(\bm r)
+H_Z(\bm B),
\label{eq:single_hole_hamiltonian}
\end{equation}
where \(H_{\mathrm{LK}}\) is the six-band Luttinger--Kohn Hamiltonian, \(H_{\mathrm{PB}}\) contains the Pikus--Bir deformation-potential terms, \(V\) is the electrostatic confinement energy, and \(H_Z\) is the linear valence-band Zeeman term.
The multiband envelope-function and strain conventions follow the standard valence-band framework~\cite{LuttingerKohn1955,Luttinger1956,BirPikus1974,Winkler2003}.
Envelope-orbital magnetic coupling enters through the covariant wave vector \(\bm{\pi}=\bm k+(e/\hbar)\bm A\).
To test whether the strain-dependent trends require explicit envelope-orbital magnetic coupling, we repeat the calculation after removing \(\bm A\) from \(\bm\pi\).
The resulting Zeeman-only magnetic-coupling model is
\begin{equation}
H_{\mathrm{Z\text{-}only}}
=H_{\mathrm{LK}}(\bm k)
+H_{\mathrm{PB}}(\bm{\epsilon})
+V(\bm r)
+H_Z(\bm B).
\label{eq:zeeman_only_hamiltonian}
\end{equation}
The explicit six-band matrix and the \(P,Q,R,S\) elements used in the solver are collected in Appendix~\ref{app:lk}.

For the six-band Hamiltonian, the device coordinates are rotated into the cubic basis
\(X,Y,Z\parallel[100],[010],[001]\):
\begin{equation}
\bm v_{\mathrm{crys}}=\bm R_{cd}\bm v_{\mathrm{dev}},
\qquad
\bm R_{cd}=
\begin{pmatrix}
1/\sqrt2&0&1/\sqrt2\\
1/\sqrt2&0&-1/\sqrt2\\
0&1&0
\end{pmatrix},
\label{eq:device_crystal_rotation}
\end{equation}
and therefore
\(\bm\epsilon_{\mathrm{crys}}=\bm R_{cd}\bm\epsilon_{\mathrm{dev}}\bm R_{cd}^{\mathsf T}\).
The resulting device-frame Pikus--Bir channels follow from this rotation and are listed in Appendix~\ref{app:lk}.
After finite-element discretization, the lowest single-hole eigenpairs
\(\{\varepsilon_i,\psi_i^\sigma(\bm r)\}\) are passed to the two-hole calculation.

\subsection{Configuration interaction}
\label{sec:ci}

The interacting two-hole Hamiltonian is
\begin{equation}
H_{\mathrm{CI}}
=\sum_i\varepsilon_i c_i^\dagger c_i
+\frac12\sum_{mnpq}W_{mnpq}
c_m^\dagger c_n^\dagger c_q c_p ,
\label{eq:ci_hamiltonian}
\end{equation}
where each index includes the six-component spinor structure.
The Coulomb matrix elements are
\begin{multline}
W_{mnpq}
=\sum_{\sigma\sigma'}
\int\dd\bm r_1\dd\bm r_2\,
\psi_m^{\sigma *}(\bm r_1)
\psi_n^{\sigma' *}(\bm r_2)\\
\times V_C(\bm r_1,\bm r_2)
\psi_p^\sigma(\bm r_1)
\psi_q^{\sigma'}(\bm r_2),
\label{eq:coulomb_matrix}
\end{multline}
with \(V_C\) obtained from the Poisson operator for the dielectric profile of the device.
The Hamiltonian is diagonalized in the antisymmetrized two-hole basis
\begin{equation}
\begin{aligned}
\Phi_{mn}(1,2)
={}&\frac{1}{\sqrt2}
\left[
\psi_m(1)\psi_n(2)\right.\\
&\left.-\psi_n(1)\psi_m(2)
\right],
\qquad m<n .
\end{aligned}
\label{eq:slater}
\end{equation}
The CI basis contains all antisymmetrized pairs formed from the retained single-hole spinors.
This full-CI construction follows the standard exact-diagonalization approach for few-particle quantum dots and retains correlations relevant to asymmetric double dots~\cite{Rontani2006,AbadilloUriel2021}.

\subsection{Fermi--Hubbard reduction and magnetic-response definitions}
\label{sec:effective_model}

The four lowest CI branches are represented by the extended two-site Fermi--Hubbard Hamiltonian~\cite{Hubbard1963,BurkardLossDiVincenzo1999}:
\begin{align}
H_{\mathrm{FH}}
={}&H_{\mathrm{ch}}
+\sum_{ss'}
\left[
c_{Ls}^{\dagger}T_{ss'}c_{Rs'}
+\mathrm{H.c.}
\right]\nonumber\\
&+\frac12\sum_{i=L,R}
c_i^\dagger
\left(\bm{\sigma}\cdot\bm b_i\right)c_i ,
\label{eq:two_site_model}
\end{align}
Here \(H_{\mathrm{ch}}=\sum_{i=L,R}(\varepsilon_i n_i+U_i n_{i\uparrow}n_{i\downarrow})+U_{LR}n_Ln_R\) contains the site energies, on-site charging costs, and interdot charging cost.
The difference of the site energies defines the detuning.
The tunneling matrix is
\begin{equation}
\begin{aligned}
\bm T
&=t_c\bm I-i t_{\mathrm{so}}\bm n_{\mathrm{so}}\cdot\bm\sigma\\
&=t\,e^{-i\theta_{\mathrm{SO}}\bm n_{\mathrm{so}}\cdot\bm\sigma},\\
\theta_{\mathrm{SO}}
&=\arctan\!\left(\frac{t_{\mathrm{so}}}{t_c}\right).
\end{aligned}
\label{eq:tunneling_angle}
\end{equation}
where \(t=(t_c^2+t_{\mathrm{so}}^2)^{1/2}\), \(t_c\) and \(t_{\mathrm{so}}\) are the spin-conserving and spin-flip amplitudes, and the unit vector \(\bm n_{\mathrm{so}}\) is their spin--orbit axis~\cite{Hetenyi2020,Geyer2024,Rodriguez2025}.
Near the selected charge anticrossing, we retain one doubly occupied singlet-like configuration together with the four predominantly separated-charge spin states.

The local Zeeman field in Eq.~\eqref{eq:two_site_model} is
\begin{equation}
\bm b_i=\mub\bm G_i\bm B,
\label{eq:local_g_field}
\end{equation}
where \(\bm G_i\) is the local magnetic-response tensor.
The mean and differential local fields are
\begin{equation}
\bar{\bm b}
=\frac{\mub}{2}\left(\bm G_L+\bm G_R\right)\bm B,
\qquad
\delta\bm b
=\frac{\mub}{2}\left(\bm G_L-\bm G_R\right)\bm B,
\label{eq:mean_difference_fields}
\end{equation}
Appendix~\ref{app:effective} evaluates the five-state matrix and derives its Schrieffer--Wolff reduction~\cite{SchriefferWolff1966}.
In the resulting singlet--triplet representation, \(\bar{\bm b}\) acts within the triplet sector, whereas \(\delta\bm b\) couples the separated-charge singlet to the triplets.
In the separated-charge product basis, the exchange sector used for the spectral fits is parameterized as
\begin{equation}
H_{\mathrm{ex}}
=\frac14\bm\sigma_L^{\mathsf T}\bm J\bm\sigma_R,
\qquad
\bm J=J_0\bm R(\bm n_{\mathrm{so}},-2\theta_{\mathrm{SO}}),
\label{eq:exchange_rotation}
\end{equation}
Here \(\bm R(\bm n,\varphi)\) is the right-handed active rotation by \(\varphi\) about \(\bm n\), and \(J_0\) is the amplitude of the fitted exchange tensor.
With the site and tunneling convention of Eq.~\eqref{eq:tunneling_angle}, the corresponding lab-frame exchange matrix carries the spin-vector rotation \(-2\theta_{\mathrm{SO}}\).
Appendix~\ref{app:effective} derives the exact five-state gap and connects these fitted coefficients to \(t_c\), \(t_{\mathrm{so}}\), and the charge-excitation energy in the perturbative charge-elimination limit.

\begin{figure}[!t]
  \centering
  \includegraphics[width=\columnwidth]{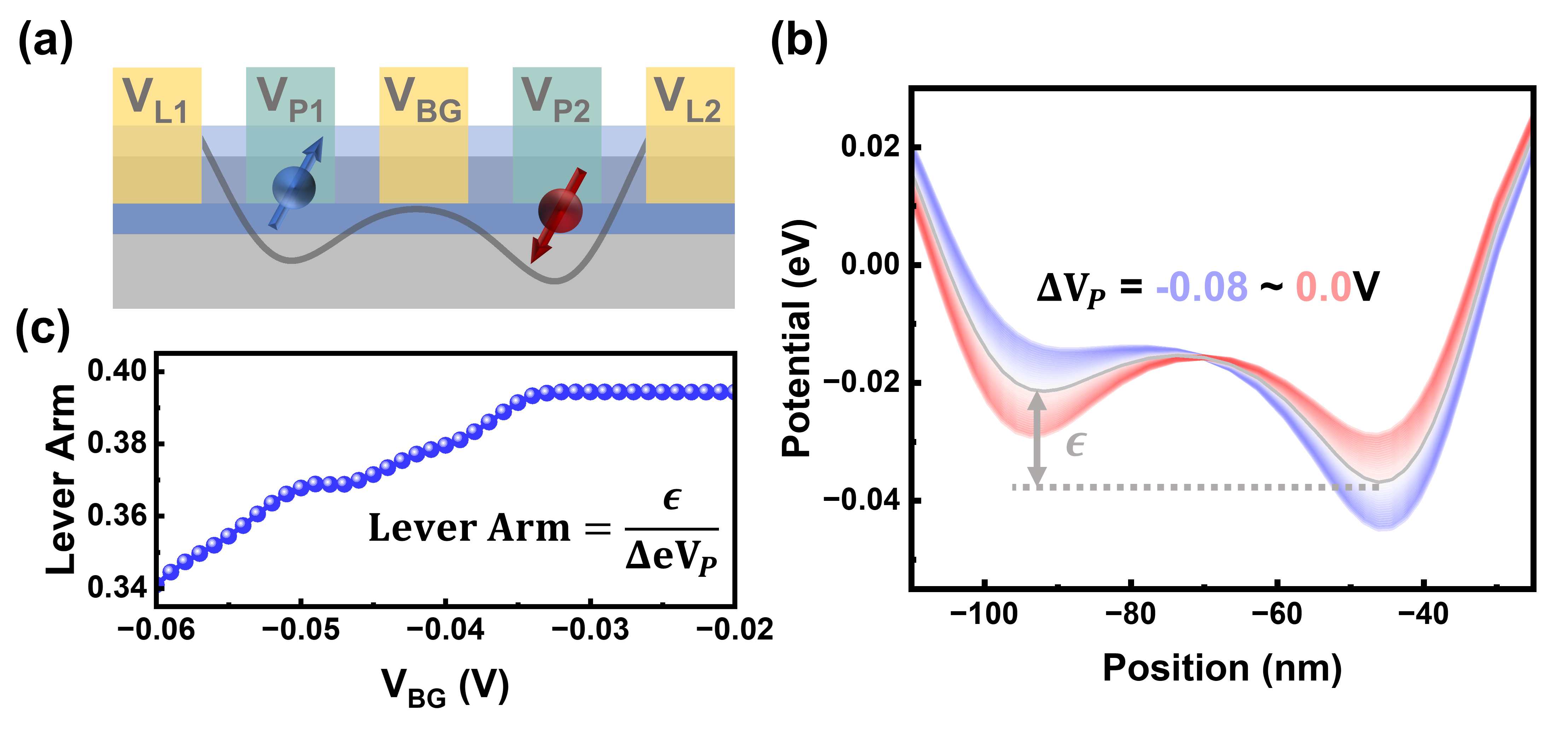}
  \caption{\label{fig:potential}
  Electrostatic detuning calibration.
  (a) Gate and double-well schematic.
  (b) Calculated longitudinal confinement potential for the plunger-voltage differences indicated by the color scale.
  The quantity \(\varepsilon\) is the energy detuning of the two minima.
  (c) Lever-arm magnitude extracted from the potential profiles as a function of barrier-gate voltage.}
\end{figure}

For each local Kramers doublet, the microscopic \(g\)-matrix \(\widetilde{\bm G}_i\) is defined in a chosen Kramers basis and changes under a rotation of that basis.
We therefore characterize the local magnetic response by the basis-invariant quadratic tensor~\cite{Crippa2018,Venitucci2018}
\begin{equation}
\bm Q_i=\widetilde{\bm G}_i^{\mathsf T}\widetilde{\bm G}_i,
\qquad
g_i(\hat{\bm n})
=\sqrt{\hat{\bm n}^{\mathsf T}\bm Q_i\hat{\bm n}}.
\label{eq:q_tensor}
\end{equation}
Its eigenvalues and eigenvectors give the squared principal \(g\) values and the corresponding principal magnetic-field directions.
The tensor \(\bm G_i\) in Eq.~\eqref{eq:local_g_field} is taken as the symmetric positive-definite square root
\begin{equation}
\bm G_i\equiv\bm Q_i^{1/2},
\qquad
\bm G_i=\bm G_i^{\mathsf T}>0 .
\label{eq:canonical_g}
\end{equation}

\section{Electrostatic calibration and exchange working point}
\label{sec:electrostatic}

The calculation with \(\bm\epsilon(\bm r)=0\) provides the unstrained reference.
The electrostatic calibration in this section and the magnetic response in Sec.~\ref{sec:baseline_magnetism} establish the baseline for the strain analysis in Sec.~\ref{sec:strain_results}.
Figures~\ref{fig:device}(a) and \ref{fig:device}(b) show the two gate-defined minima and the complementary roles of the controls: the plunger pair tilts the double well, whereas \(BG\) changes the barrier between the dots.
The resulting barrier control of the coupling is quantified in Fig.~\ref{fig:device}(c).
Raising the barrier suppresses both \(t_c\) and \(J\); the stronger variation of \(J\) is consistent with the \(t_c^2\)-type kinetic-exchange scaling generated by virtual interdot tunneling in the Hubbard limit~\cite{LossDiVincenzo1998,BurkardLossDiVincenzo1999,Burkard2023}.
The sweep therefore establishes an electrically accessible exchange range.
The specific operating point is selected only after the charge spectrum has been resolved.

Opposite shifts of the two plunger voltages define the detuning coordinate.
The calculated potential profiles in Fig.~\ref{fig:potential}(b) convert this voltage displacement into the energy difference \(\varepsilon\) between the two minima.
The corresponding lever arm \(\alpha=|\varepsilon/(e\Delta V_P)|\) varies with the barrier setting before approaching a plateau [Fig.~\ref{fig:potential}(c)].
Accounting for this variation places the spectra at different barrier voltages on a common energy-detuning scale.

\begin{figure}[!t]
  \centering
  \includegraphics[width=\columnwidth]{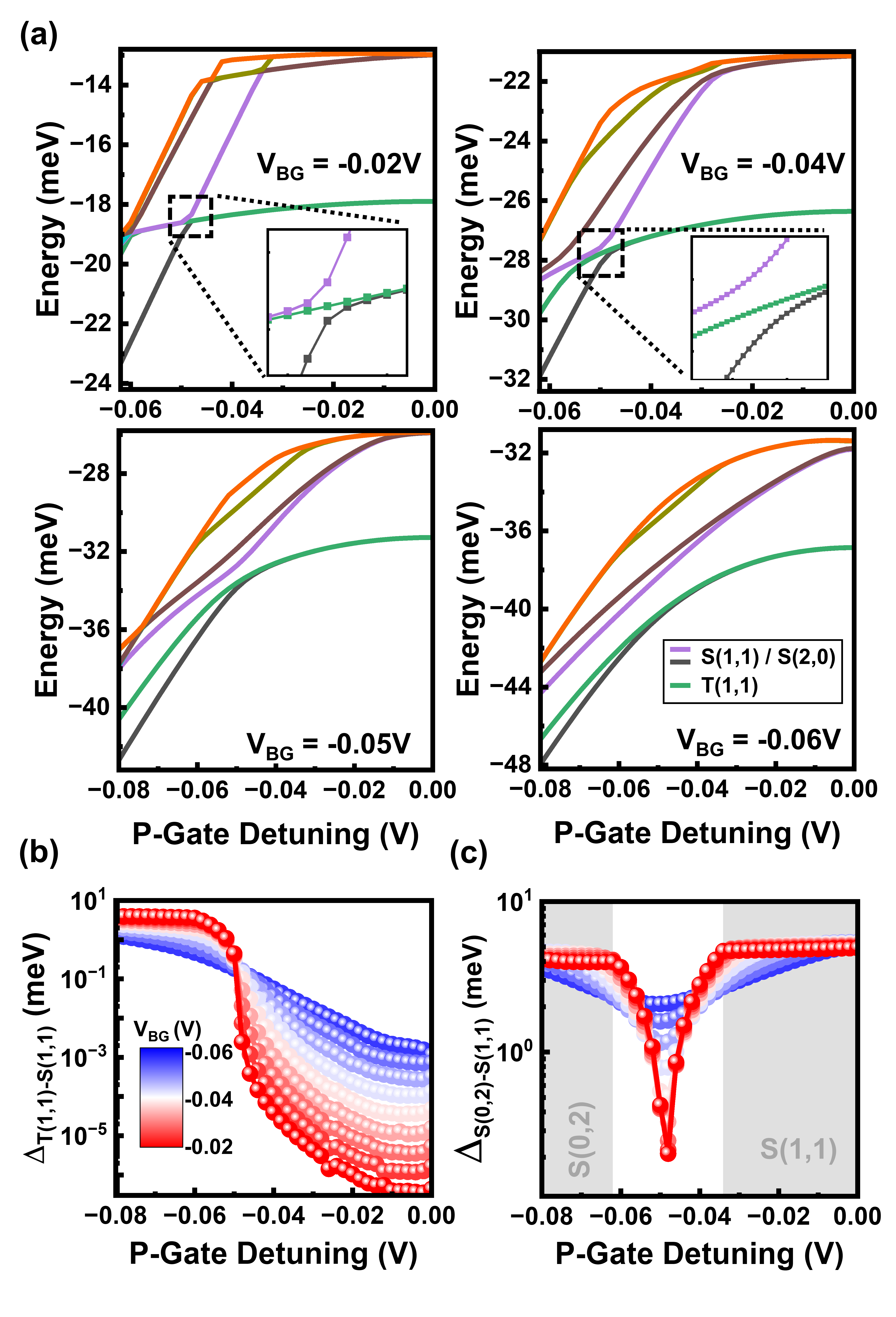}
  \caption{\label{fig:detuning}
  CI two-hole spectra and energy gaps under plunger detuning.
  Panel (a) shows the low-energy branches for the four indicated barrier-gate voltages.
  Panels (b) and (c) summarize the singlet-like charge anticrossing and the separation of the low-energy spin branches.
  Gray regions mark the predominantly doubly occupied and separated-charge regimes.
  Energies are plotted relative to the reference used in each numerical sweep.}
\end{figure}

Figure~\ref{fig:detuning} then follows the correlated two-hole spectrum through the charge anticrossing.
For every barrier setting, the two singlet-like branches exchange their dominant separated-charge and doubly occupied character, while the lowest spin manifold remains on a much smaller energy scale.
The insets in Fig.~\ref{fig:detuning}(a) resolve the low-energy spin branches over the same detuning interval.
Varying \(V_{\mathrm{BG}}\) changes both the curvature of the singlet-like branches and the residual splitting within the spin manifold, showing that charge hybridization and the low-energy spin scale must be calibrated together.

\begin{figure}[!t]
  \centering
  \includegraphics[width=\columnwidth]{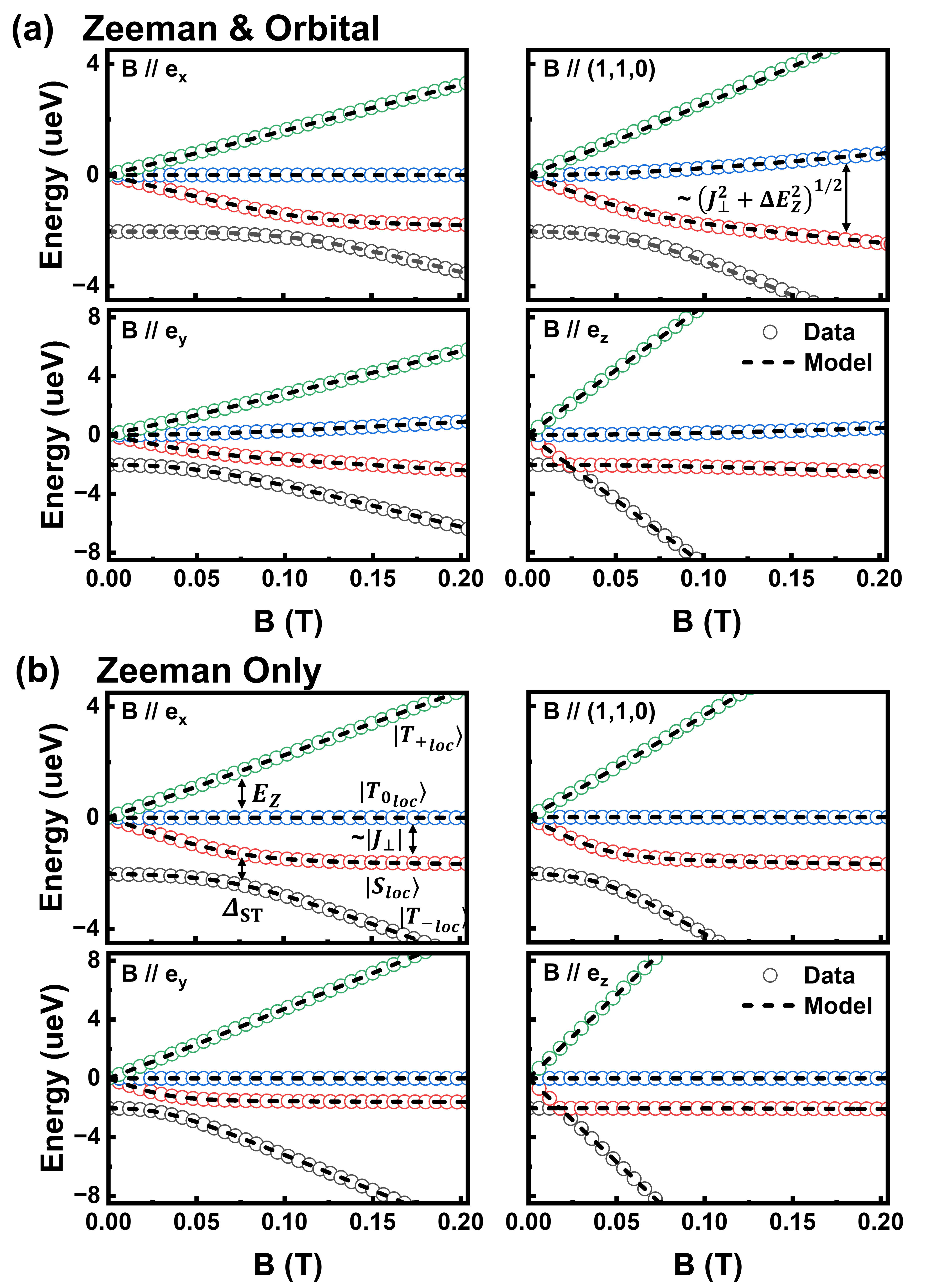}
  \caption{\label{fig:magnetic_spectrum}
  Unstrained directional two-hole spectra.
  (a) Full calculation including both Zeeman and envelope-orbital magnetic couplings.
  (b) Zeeman-only calculation.
  Within each group the field is oriented along \(\bm e_x\), \([110]\), \(\bm e_y\), and \(\bm e_z\).
  Open circles denote CI eigenenergies and dashed curves denote the fitted low-energy-model branches.
  The annotations identify the Zeeman spacing \(E_Z\), the local splitting mismatch \(\Delta E_Z\), the transverse-exchange splitting \(J_\perp\), and the singlet--triplet avoided-crossing gap \(\Delta_{\mathrm{ST}}\).}
\end{figure}

\begin{figure}[!t]
  \centering
  \includegraphics[width=\columnwidth]{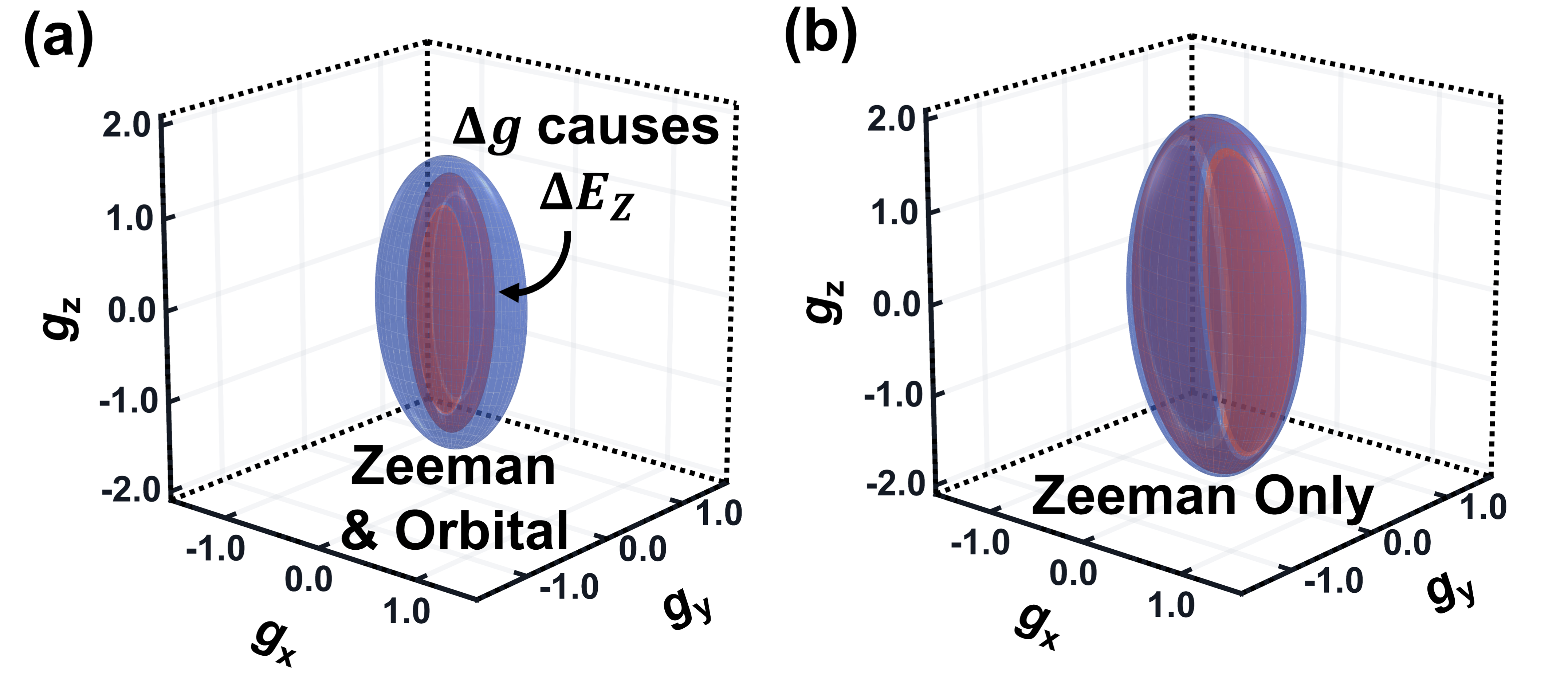}
  \caption{\label{fig:unstrained_ellipsoids}
  Unstrained field-space response ellipsoids obtained from the direction-resolved spectrum fits for (a) the full Hamiltonian and (b) the Zeeman-only Hamiltonian.
  Red and blue identify the two fitted local tensors represented by the symmetric positive-definite matrices \(\bm G_i=\bm Q_i^{1/2}\).
  Their semiaxes and orientations give the principal response values and principal magnetic-field directions, respectively.
  The arrow in panel (a) emphasizes that a difference between the local responses produces the splitting mismatch \(\Delta E_Z\) in Eq.~\eqref{eq:local_zeeman_mismatch}.}
\end{figure}

Figures~\ref{fig:detuning}(b) and \ref{fig:detuning}(c) make this separation of energy scales explicit.
The charge gap is minimized near the anticrossing, whereas the spin gap continues to evolve across the separated-charge regime.
The colored traces further show that this low-energy scale remains sensitive to the barrier voltage away from the center of the anticrossing.
These trends identify a detuning window in which the charge-excited singlet remains well isolated while the four lowest spin states form the reduced manifold used below.

Guided by these sweeps, an additional calculation at \(V_{\mathrm{BG}}=-0.035~\mathrm V\) and \(\Delta V_P=-0.042~\mathrm V\) defines the common working point for the magnetic analysis.
There the charge-excited singlet is separated by \(1.900~\mathrm{meV}\), while the lowest singlet--triplet splitting is only \(J=2.031~\ueV\) (\(J/h=491~\mathrm{MHz}\)).
This three-order-of-magnitude separation justifies retaining the four lowest spin states.

\section{Magnetic spectra and tensor baseline}
\label{sec:baseline_magnetism}

With the charge manifold fixed, Fig.~\ref{fig:magnetic_spectrum} compares the unstrained magnetic spectrum from the full calculation with its Zeeman-only counterpart for four field directions.

In the full spectrum [Fig.~\ref{fig:magnetic_spectrum}(a)], two branches disperse outward while the other two remain near the center.
Away from the avoided crossings, the outer-branch dispersion follows the mean local response
\(\bar g(\hat{\bm n})=[g_L(\hat{\bm n})+g_R(\hat{\bm n})]/2\).
This accounts for the strong dispersion along \(z\) and the much weaker dispersion along \(x\).
The central pair also senses the difference between the two local Zeeman splittings,
\begin{equation}
\Delta E_Z(\hat{\bm n})
=\mub B\left[g_L(\hat{\bm n})-g_R(\hat{\bm n})\right].
\label{eq:local_zeeman_mismatch}
\end{equation}
The transverse-exchange splitting in the local quantization axes set by the two Zeeman fields is
\begin{equation}
J_\perp(\hat{\bm n})
=2\left|
\bra{\uparrow_L\downarrow_R}
H_{\mathrm{ex}}
\ket{\downarrow_L\uparrow_R}
\right|.
\label{eq:transverse_exchange_splitting}
\end{equation}
The factor of two converts the off-diagonal matrix element into the level splitting of the degenerate antiparallel-spin pair.
The leading separation of the central pair is then
\(\Delta E_{\mathrm{inner}}\simeq\sqrt{J_\perp^2+\Delta E_Z^2}\), as indicated for \(\bm B\parallel[110]\).

The Zeeman-only spectra in Fig.~\ref{fig:magnetic_spectrum}(b) make the roles of these two terms particularly clear.
Here the two local responses are nearly equal, so \(\Delta E_Z\) is small for all four directions and the inner branches remain almost parallel.
Away from the avoided crossing, the branches may consequently be identified with the local-basis states
\(\ket{T_{+,\mathrm{loc}}}\), \(\ket{T_{0,\mathrm{loc}}}\), \(\ket{S_{\mathrm{loc}}}\), and \(\ket{T_{-,\mathrm{loc}}}\).
The mean response gives the marked Zeeman spacing \(E_Z\), while the \(\ket{S_{\mathrm{loc}}}\)--\(\ket{T_{0,\mathrm{loc}}}\) separation reduces to \(\lvert J_\perp\rvert\).

\begin{figure}[!t]
  \centering
  \includegraphics[width=0.78\columnwidth]{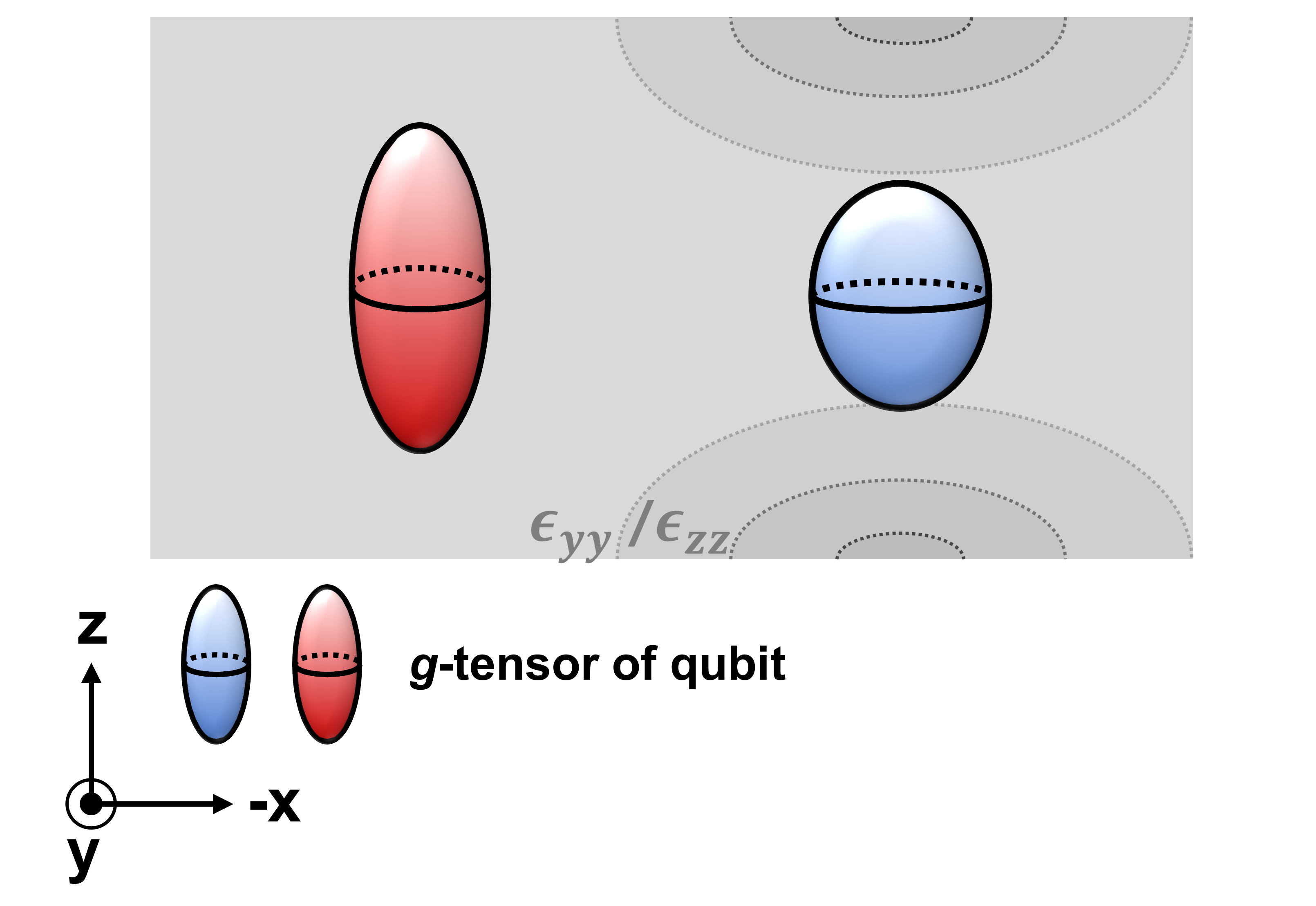}
  \caption{\label{fig:diagonal_strain_schematic}
  Spatial protocol for differential diagonal strain.
  A device-frame \(\epsilon_{yy}\) or \(\epsilon_{zz}\) profile is applied to the right dot, with the left dot providing the unstrained reference.
  Red and blue denote the two local magnetic-response tensors.}
\end{figure}

\begin{figure}[!t]
  \centering
  \includegraphics[width=\columnwidth]{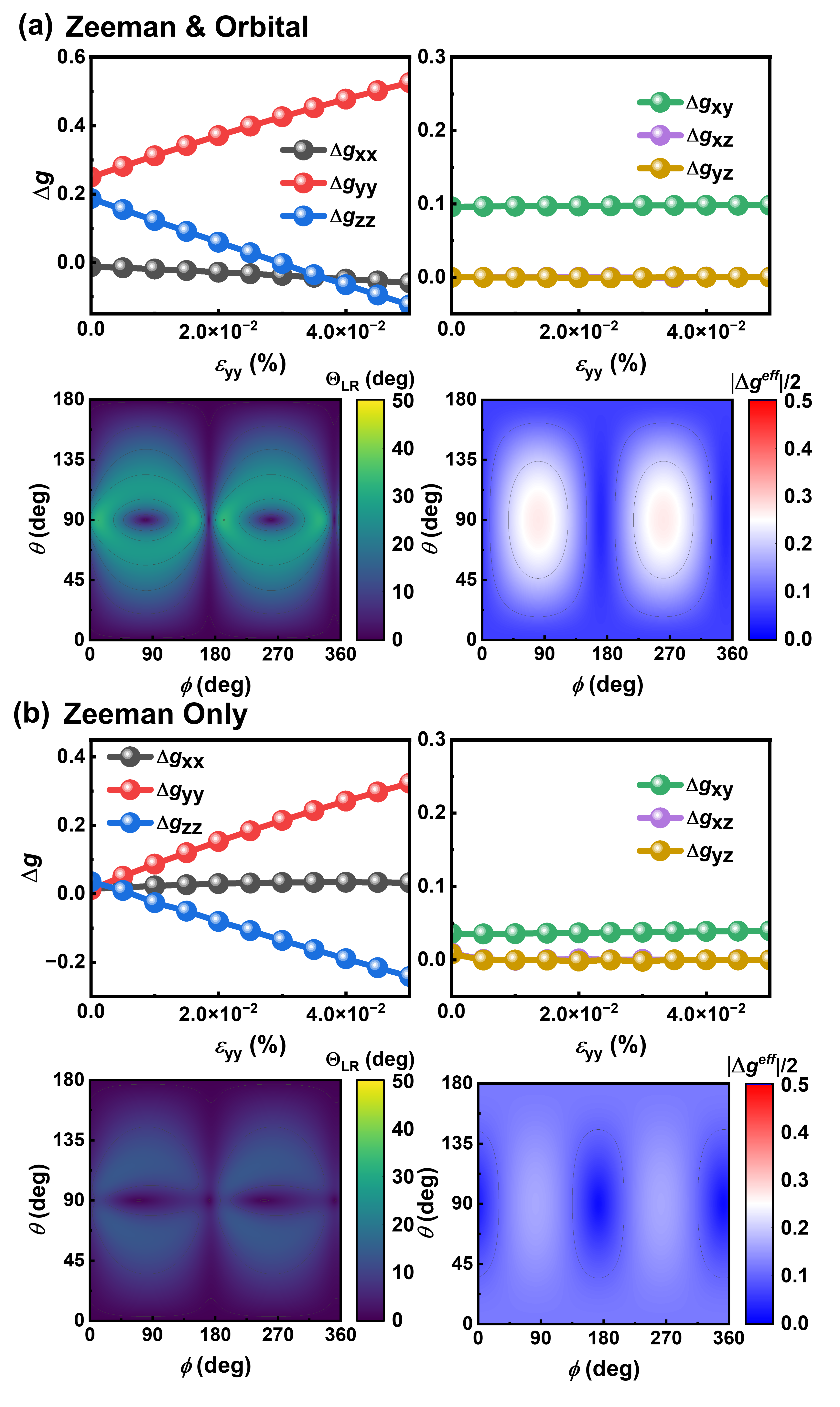}
  \caption{\label{fig:strain_yy}
  Magnetic-response mismatch induced by differential \(\epsilon_{yy}\) strain.
  Rows (a) and (b) show the full (Zeeman and orbital) and Zeeman-only calculations, respectively.
  In each row, the upper-left and upper-right panels give the diagonal and off-diagonal laboratory-frame elements \(\Delta g_{\alpha\beta}=[\Delta\bm G]_{\alpha\beta}\) of the continuity-aligned tensor difference \(\Delta\bm G=\bm G_1-\bm G_2\).
  At \(\epsilon_{yy}=0.050\%\), the lower-left and lower-right panels map the relative canonical-vector angle \(\Theta_{LR}\) and the vector half mismatch \(\left|\Delta\bm g^{\mathrm{eff}}\right|/2\), respectively.
  Here \(\theta\) is measured from the laboratory \(z\) axis and \(\phi\) in the \(x\)--\(y\) plane.}
\end{figure}

\begin{figure}[!t]
  \centering
  \includegraphics[width=\columnwidth]{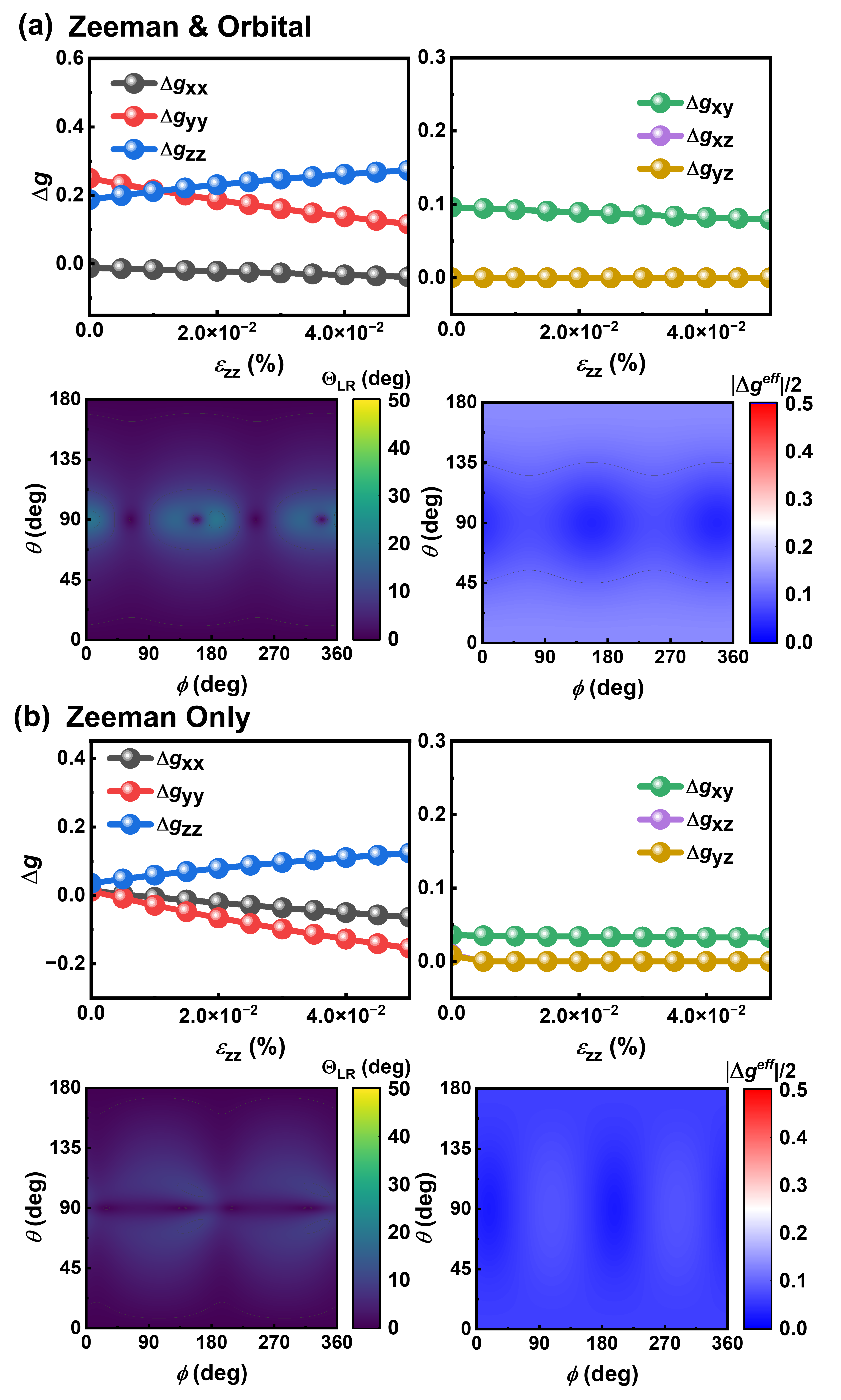}
  \caption{\label{fig:strain_zz}
  Magnetic-response mismatch induced by differential \(\epsilon_{zz}\) strain.
  Rows (a) and (b) show the full (Zeeman and orbital) and Zeeman-only calculations, respectively.
  In each row, the upper-left and upper-right panels give the diagonal and off-diagonal elements of \(\Delta\bm G\).
  At \(\epsilon_{zz}=0.050\%\), the lower-left panel maps \(\Theta_{LR}\), and the lower-right panel maps \(\left|\Delta\bm g^{\mathrm{eff}}\right|/2\).
  These angular panels use the same scales and canonical square-root convention as Fig.~\ref{fig:strain_yy}.
  Their weaker contrast shows that the directional mismatch generated by \(\epsilon_{zz}\) is smaller.}
\end{figure}

\begin{figure}[!t]
  \centering
  \includegraphics[width=\columnwidth]{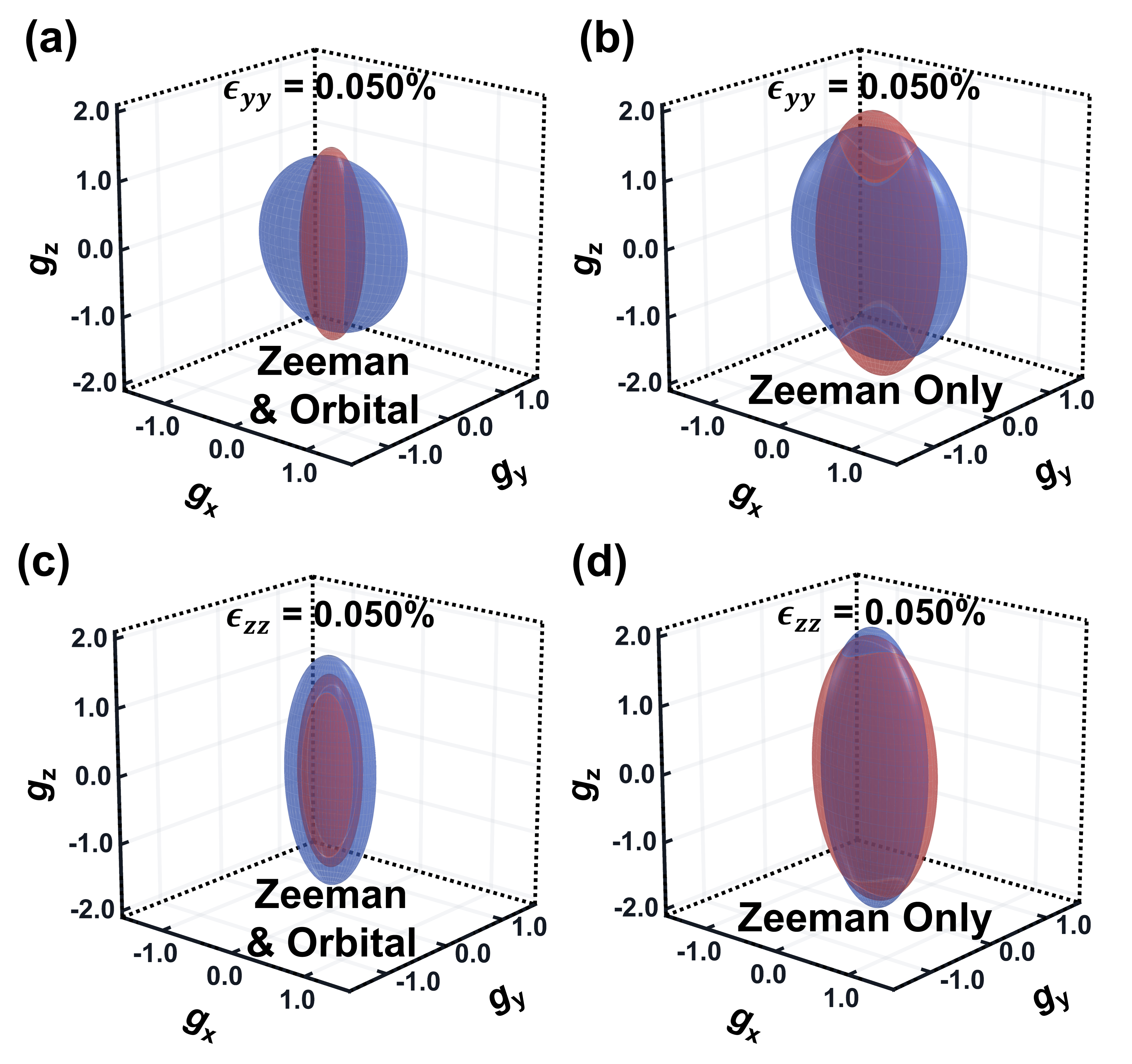}
  \caption{\label{fig:diagonal_ellipsoids}
  Endpoint field-space response ellipsoids under differential diagonal strain.
  Panels (a) and (b) show \(\epsilon_{yy}=0.050\%\) for the full and Zeeman-only Hamiltonians, respectively.
  Panels (c) and (d) show the corresponding results for \(\epsilon_{zz}=0.050\%\).
  Red and blue identify the two local tensors in the same symmetric-positive-definite representation \(\bm G_i=\bm Q_i^{1/2}\) as in Fig.~\ref{fig:unstrained_ellipsoids}.
  The semiaxis changes summarize the component-dependent redistribution of the principal \(g\) values, while the nearly common longest-axis orientations show the small opening of the maximum-response axes.}
\end{figure}

\begin{figure}[!t]
  \centering
  \includegraphics[width=0.78\columnwidth]{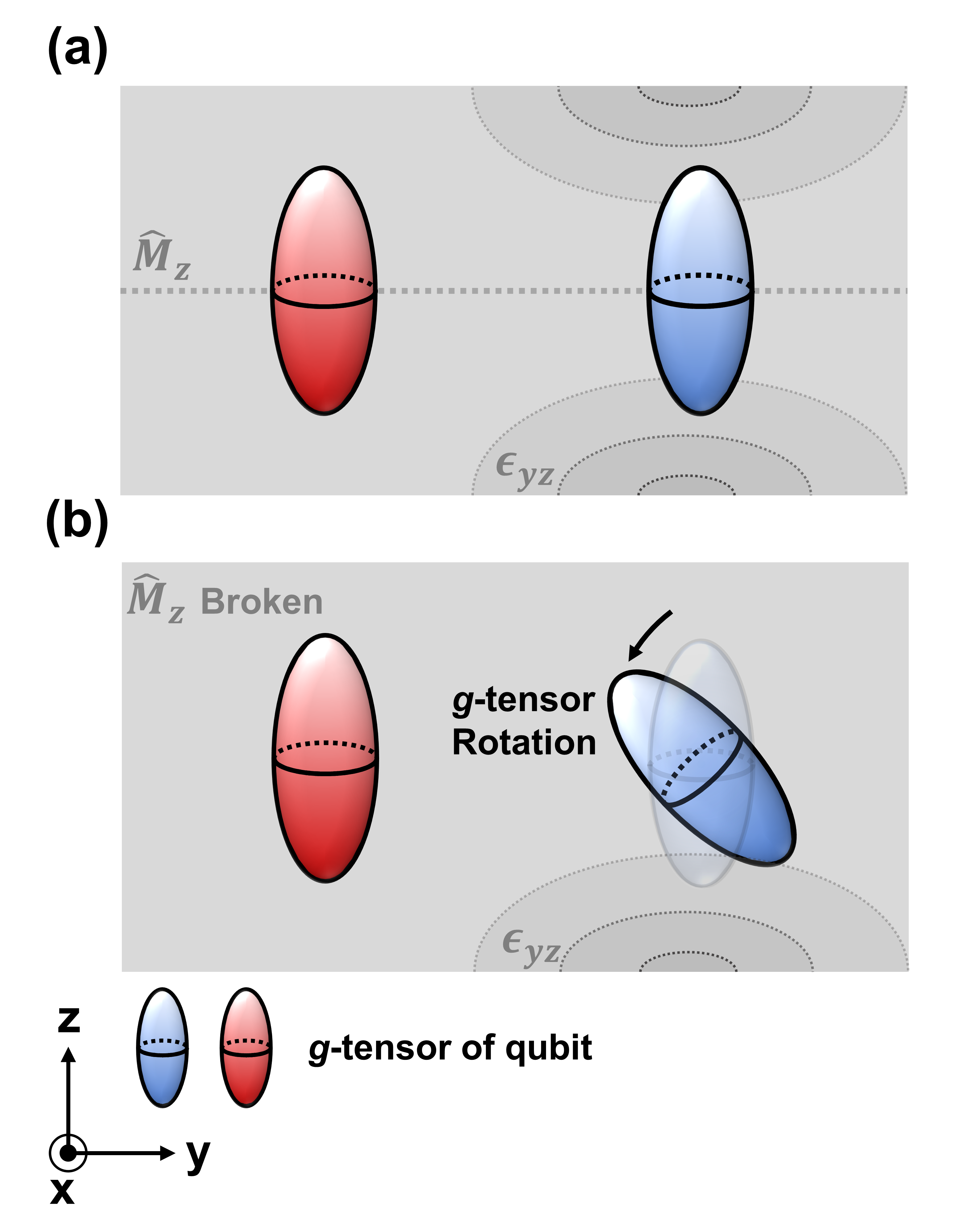}
  \caption{\label{fig:offdiagonal_strain_schematic}
  Spatial-profile comparison for off-diagonal \(\epsilon_{yz}\).
  (a) The \(z\)-odd profile is compatible with the transverse device mirror \(M_z\).
  (b) The \(z\)-even profile breaks this local constraint and permits the response ellipsoids to rotate.
  Red and blue schematically represent the two local response tensors, and the dotted lines indicate their principal axes.
  The two profiles have the same strain-tensor component and maximum magnitude.}
\end{figure}

At the lower singlet--triplet encounter, the couplings \(\beta_0\) and \(\beta_\pm\) in Eq.~\eqref{eq:mixed_couplings} open the gap \(\Delta_{\mathrm{ST}}\).
They contain both the differential local field and the spin--orbit tunneling vector \(t_{\mathrm{so}}\bm n_{\mathrm{so}}\)~\cite{Hwang2017,Tanttu2019,Geyer2024,Rodriguez2025}.

The dashed branches are least-squares fits of the four-state model to the CI eigenenergies.
The global energy root-mean-square errors are \(2.36\times10^{-2}~\ueV\) for the full calculation and \(6.43\times10^{-3}~\ueV\) for the Zeeman-only calculation.
The fitted model thus resolves both the anisotropic outer dispersion and the exchange- and mismatch-controlled inner structure.

Figure~\ref{fig:unstrained_ellipsoids} makes the tensor origin of the spectral contrast explicit.
In the full calculation, the fitted local ellipsoids differ in their transverse semiaxes and principal directions.
This produces a direction-dependent \(\Delta E_Z\) and accounts for the additional inner-branch separation in Fig.~\ref{fig:magnetic_spectrum}(a).
The Zeeman-only ellipsoids are nearly coincident, consistent with \(\Delta E_Z\simeq0\), an inner separation dominated by \(J_\perp\), and the nearly parallel central branches in Fig.~\ref{fig:magnetic_spectrum}(b).
Both tensor pairs have their longest principal axes close to \(z\), which explains why the outer branches disperse most strongly for \(\bm B\parallel\bm e_z\).
They provide the respective unstrained references for the strain-dependent calculations below.

\section{Symmetry-selective strain control}
\label{sec:strain_results}

At the electrostatic working point established above, the strain in the right-dot region is varied while the left-dot region remains unstrained.
This protocol creates a local strain difference without changing the gate-defined confinement.
Thermal strain in a FinFET can contain a longitudinal component \(\epsilon_{xx}\) together with \(\epsilon_{yy}\), \(\epsilon_{zz}\), and \(\epsilon_{yz}\)~\cite{Bouquet2025}.
Because the dots are arranged along the mechanically continuous fin axis, a controlled dot-to-dot difference in \(\epsilon_{xx}\) would require an axial strain gradient on the interdot length scale.
We focus on \(\epsilon_{yy}\), \(\epsilon_{zz}\), and \(\epsilon_{yz}\), for which differential profiles can be imposed directly in the present geometry.
For each profile, the full calculation is paired with a Zeeman-only calculation that shows how envelope-orbital magnetic coupling changes the response.

\subsection{Diagonal strain: component-selective reshaping}
\label{sec:diagonal_strain}

We first compare \(\epsilon_{yy}\) and \(\epsilon_{zz}\) using the common spatial protocol in Fig.~\ref{fig:diagonal_strain_schematic}.
Each component is swept from \(0\) to \(0.050\%\) in steps of \(0.005\%\).
At every strain point, the direction-resolved CI eigenenergies are fitted by least squares to the four-state two-spin model of Sec.~\ref{sec:effective_model}, and the plotted \(\Delta\bm G\) is formed from the two fitted local tensors.
Across all fits used in Figs.~\ref{fig:strain_yy} and \ref{fig:strain_zz}, the largest root-mean-square energy residual is \(6.2\times10^{-3}~\ueV\).
The increment
\(\delta_\epsilon\Delta\bm G=\Delta\bm G(\epsilon)-\Delta\bm G(0)\)
removes the different unstrained backgrounds from the strain comparison.

The \(\epsilon_{yy}\) sweep produces an almost entirely diagonal change of the local-tensor difference [Fig.~\ref{fig:strain_yy}].
Both magnetic-coupling models show the same component selection: \(\Delta g_{yy}\) increases, \(\Delta g_{zz}\) decreases through zero, and the off-diagonal elements vary only weakly.
At \(\epsilon_{yy}=0.050\%\), the Frobenius norms of the strain-induced increments are \(0.418\) and \(0.417\) for the full and Zeeman-only calculations, respectively.
Their nearly equal magnitudes conceal a redistribution among the diagonal components: the full calculation gives
\((\delta_\epsilon\Delta g_{yy},\delta_\epsilon\Delta g_{zz})=(0.276,-0.311)\), compared with \((0.311,-0.277)\) in the Zeeman-only calculation, and the smaller \(xx\) increment also changes sign.

The lower panels of Figs.~\ref{fig:strain_yy}(a) and \ref{fig:strain_yy}(b) resolve the endpoint response over the magnetic-field direction.
For \(\hat{\bm n}=\bm B/B\), the canonical response vectors, their difference, and their relative angle are
\begin{align}
\bm g_i^{\mathrm{eff}}(\hat{\bm n})
&=\bm G_i\hat{\bm n},
\qquad
\Delta\bm g^{\mathrm{eff}}
=\bm g_L^{\mathrm{eff}}-\bm g_R^{\mathrm{eff}},
\label{eq:directional_tensor_map}\\
\Theta_{LR}(\hat{\bm n})
&=\cos^{-1}
\frac{
\bm g_L^{\mathrm{eff}}\cdot\bm g_R^{\mathrm{eff}}
}{
\left|\bm g_L^{\mathrm{eff}}\right|
\left|\bm g_R^{\mathrm{eff}}\right|
}.
\label{eq:axis_angle}
\end{align}
Here \(\Theta_{LR}\) records the angle between the two vectors, whereas
\(\left|\Delta\bm g^{\mathrm{eff}}\right|/2\) records their separation in both magnitude and direction.
Both are geometric diagnostics of the canonical representatives \(\bm G_i=\bm Q_i^{1/2}\).
In Figs.~\ref{fig:strain_yy}(a) and \ref{fig:strain_yy}(b), the largest vector mismatch lies near \(y\), where \(\Theta_{LR}<1.5^\circ\).
The dominant endpoint difference is therefore an unequal response magnitude along nearly the same canonical-vector direction.
Away from \(y\), the full calculation shows the stronger angular structure, although the two maximum-response axes remain nearly coincident.

The \(\epsilon_{zz}\) sweep produces the opposite ordering of the two leading changes: \(\Delta g_{zz}\) increases and \(\Delta g_{yy}\) decreases [Fig.~\ref{fig:strain_zz}].
This trend is again common to the two magnetic-coupling models.
At \(\epsilon_{zz}=0.050\%\), the \(zz\) increment is nearly unchanged between the full and Zeeman-only calculations (\(0.086\) and \(0.088\)), whereas the accompanying \(yy\) changes are \(-0.133\) and \(-0.166\), respectively.
The full-model Frobenius increment is \(0.163\), about \(39\%\) of its \(\epsilon_{yy}\) counterpart, and the response remains predominantly diagonal in both calculations.
The angular maps are weaker than for \(\epsilon_{yy}\).
In the full calculation the largest vector mismatch lies along \(z\), where the two effective Zeeman vectors are essentially collinear, whereas the Zeeman-only maximum lies closer to the equatorial plane.
Thus \(\epsilon_{yy}\) and \(\epsilon_{zz}\) tune the magnitudes of the local responses in different directions and with markedly different strengths.

The endpoint ellipsoids in Fig.~\ref{fig:diagonal_ellipsoids} provide a principal-axis view of this component contrast.
The \(\epsilon_{yy}\) protocol contracts the longest semiaxis and expands the intermediate one, while \(\epsilon_{zz}\) slightly extends the longest semiaxis and contracts the transverse response.
The same contrasting deformation appears in the Zeeman-only calculation, showing that the valence-band Zeeman interaction is sufficient to generate the diagonal component selection.
The full and Zeeman-only calculations start from the distinct unstrained tensor pairs in Fig.~\ref{fig:unstrained_ellipsoids}.
These baseline differences remain visible in the endpoint ellipsoids of Fig.~\ref{fig:diagonal_ellipsoids} and the angular maps of Figs.~\ref{fig:strain_yy} and \ref{fig:strain_zz}.

The different trends follow the device-to-crystal mapping of the Pikus--Bir perturbation [Eq.~\eqref{eq:device_pb_channels}]~\cite{BirPikus1974,Winkler2003,Wang2024}.
Device-frame \(\epsilon_{yy}\) enters \(P_\epsilon\) and \(Q_\epsilon\), whereas \(\epsilon_{zz}\) also activates \(R_\epsilon=i d\epsilon_{zz}/2\).
The two components consequently perturb different combinations of heavy-hole, light-hole, and split-off admixtures even at equal nominal strain.

\begin{figure}[!t]
  \centering
  \includegraphics[width=\columnwidth]{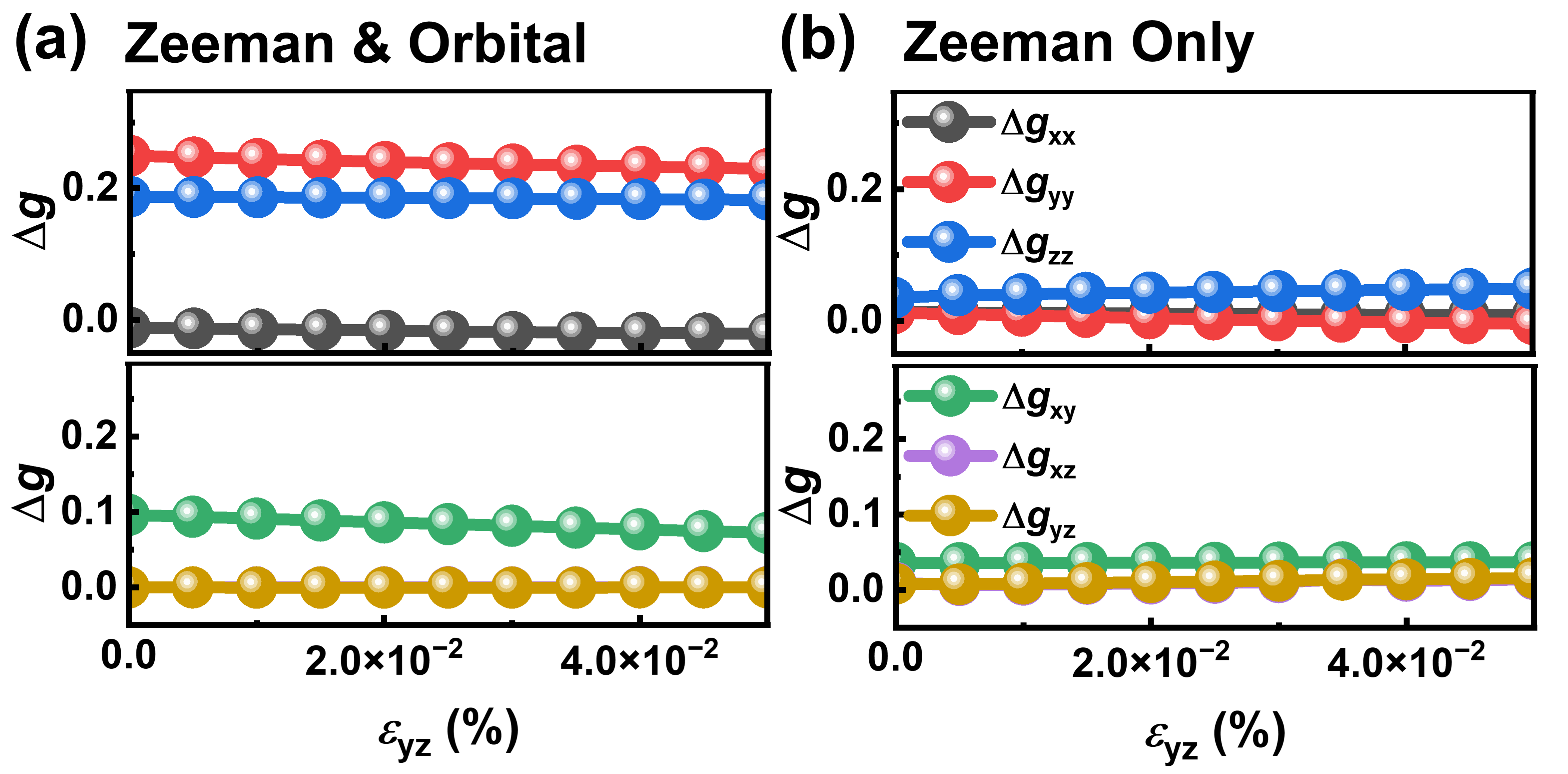}
  \caption{\label{fig:strain_yz_preserving}
  Local-tensor response to the mirror-compatible, \(z\)-odd \(\epsilon_{yz}\) profile.
  Panels (a) and (b) show the full and Zeeman-only calculations, respectively.
  Within each panel, the upper and lower plots resolve the diagonal and off-diagonal elements of \(\Delta\bm G\).
  In both calculations, all six elements remain close to their zero-strain baselines throughout the sweep.}
\end{figure}

\begin{figure}[!t]
  \centering
  \includegraphics[width=\columnwidth]{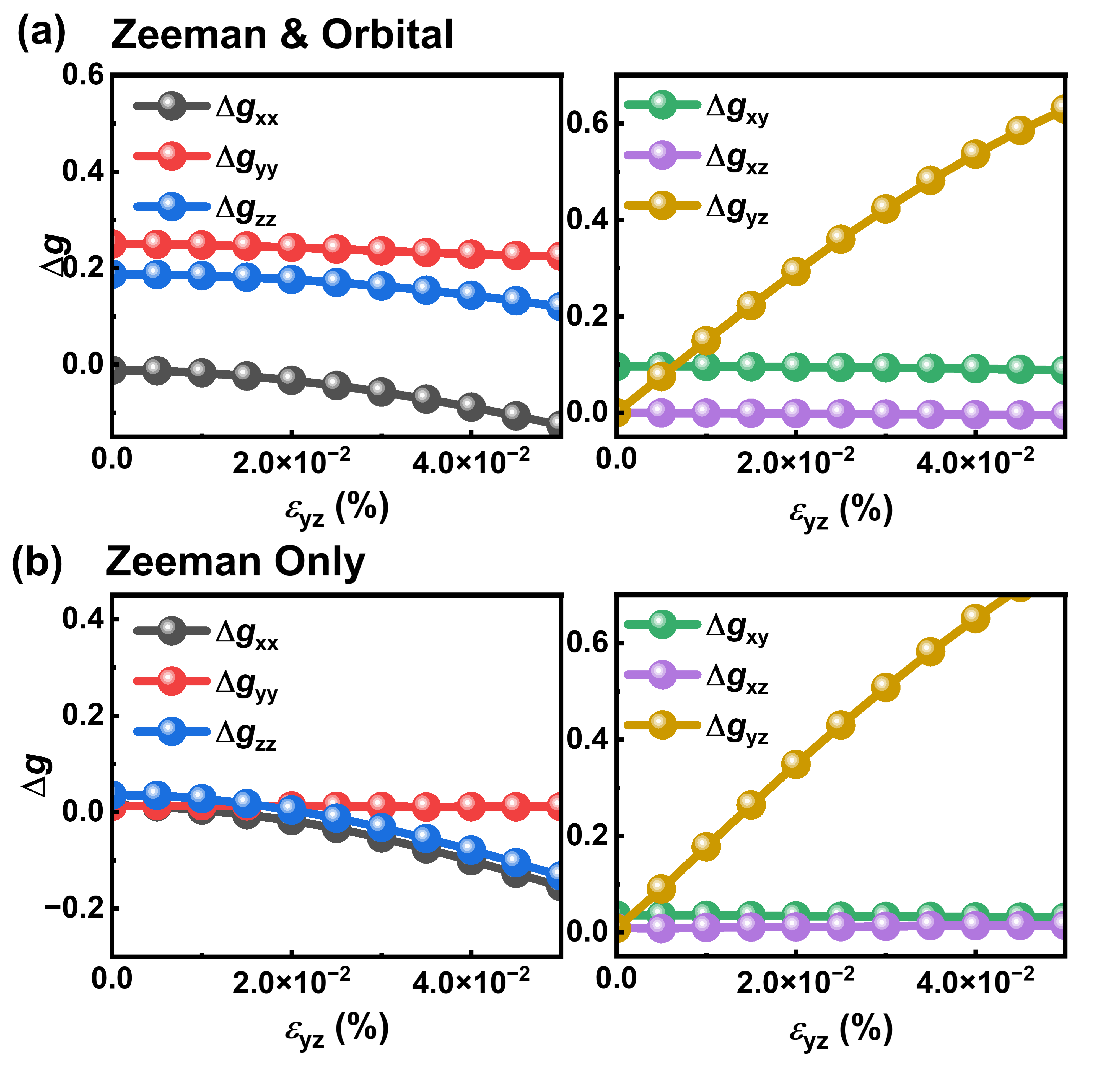}
  \caption{\label{fig:strain_yz_breaking_components}
  Tensor response under the mirror-breaking, \(z\)-even \(\epsilon_{yz}\) profile.
  The upper and lower rows show the full and Zeeman-only calculations, respectively.
  The left and right panels resolve the diagonal and off-diagonal elements of \(\Delta\bm G\).
  In both calculations, an almost linear \(\Delta g_{yz}\) becomes the dominant strain-induced component.}
\end{figure}

\subsection{Shear strain: spatial symmetry as a selection rule}
\label{sec:offdiagonal_strain}

For the shear channel, the parity across the transverse midplane determines whether the applied strain is compatible with the local mirror.
Figure~\ref{fig:offdiagonal_strain_schematic} compares a \(z\)-odd profile and a \(z\)-even profile with the same maximum \(\lvert\epsilon_{yz}\rvert\), each swept from \(0\) to \(0.050\%\) in steps of \(0.005\%\):
\begin{align*}
\epsilon_{yz}(x,y,z)&=-\epsilon_{yz}(x,y,-z)
&& (z\text{-odd}),\\
\epsilon_{yz}(x,y,z)&=+\epsilon_{yz}(x,y,-z)
&& (z\text{-even}).
\end{align*}
Under \(M_z:(x,y,z)\mapsto(x,y,-z)\), the tensor component transforms as
\(\epsilon_{yz}'(x,y,z)=-\epsilon_{yz}(x,y,-z)\).
The extra minus sign makes the \(z\)-odd profile compatible with \(M_z\), whereas a nonzero \(z\)-even profile breaks it.
This reflection leaves the dot labels unchanged and constrains the response locally at each dot.
In the crystal frame, device-frame \(\epsilon_{yz}\) enters the off-diagonal Pikus--Bir channel
\(S_\epsilon=-d(1+i)\epsilon_{yz}/\sqrt2\) [Eq.~\eqref{eq:device_pb_channels}].
The two ingredients play different roles: \(S_\epsilon\) supplies the multiband coupling associated with an off-diagonal response, while the transverse parity decides whether the local \(xz/yz\) magnetic-response elements are symmetry allowed.

We extract the local tensors with the direction-resolved least-squares procedure used for the diagonal sweeps.
Across all strain points in the four series shown in Figs.~\ref{fig:strain_yz_preserving} and \ref{fig:strain_yz_breaking_components}, the largest energy RMSE is \(3.9\times10^{-3}~\ueV\).

For the \(z\)-odd profile (\(M_z\) symmetry preserved), both the diagonal and off-diagonal elements remain close to their zero-strain baselines in the full and Zeeman-only calculations [Fig.~\ref{fig:strain_yz_preserving}].
The two Hamiltonians give different background tensors, but neither shows an appreciable strain-induced change over this sweep.
The full-model maximum-response axes remain aligned to within \(0.02^\circ\), with similarly negligible opening in the Zeeman-only calculation.
The local mirror constraint specifically forbids the \(xz/yz\) rotation block, while the numerically weak diagonal and \(xy\) changes are features of the present profile and device.

\begin{figure}[!t]
  \centering
  \includegraphics[width=\columnwidth]{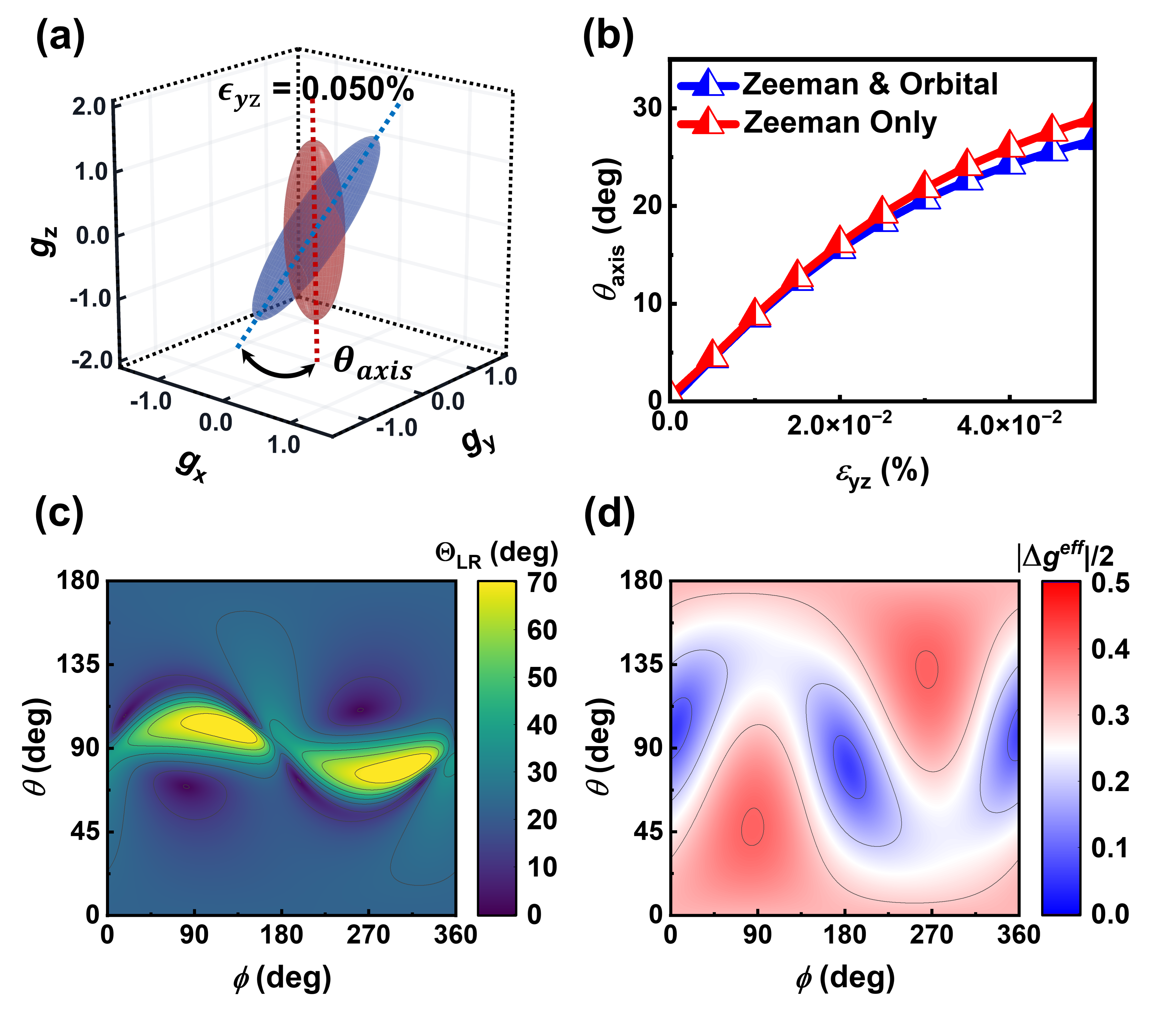}
  \caption{\label{fig:strain_yz_breaking_geometry}
  Geometric and directional signatures of the mirror-breaking, \(z\)-even \(\epsilon_{yz}\) response.
  (a) Full-model endpoint response ellipsoids at \(\epsilon_{yz}=0.050\%\).
  Red and blue identify the two local tensors, and dotted lines mark their maximum-response axes.
  (b) Continuity-tracked opening angle of their maximum-response axes.
  Blue and red denote the full and Zeeman-only calculations, respectively.
  (c) Relative canonical-vector angle \(\Theta_{LR}\) and (d) vector half mismatch \(\left|\Delta\bm g^{\mathrm{eff}}\right|/2\) for the full-model endpoint tensors \(\bm G_i\).
  The \(\Theta_{LR}\) color scale is clipped at \(70^\circ\).
  Larger values share the saturated color.
  The full/Zeeman-only comparison uses the ten common nonzero strain points.
  The zero-strain marker is shown only as a separate reference.}
\end{figure}

For the \(z\)-even profile (\(M_z\) symmetry broken), both the diagonal and off-diagonal sectors of \(\Delta\bm G\) evolve with strain [Fig.~\ref{fig:strain_yz_breaking_components}].
Alongside the diagonal and \(xy\) evolution, \(\Delta g_{yz}\) grows nearly linearly and becomes the largest strain-induced component, while \(\Delta g_{xz}\) remains much smaller.
The dominant \(yz\) element couples the \(y\) and \(z\) directions in the device basis and is accompanied by a pronounced principal-axis rotation.
At \(\epsilon_{yz}=0.050\%\), the maximum-response axes open by about \(27^\circ\) in the full calculation and \(29^\circ\) in the Zeeman-only calculation [Fig.~\ref{fig:strain_yz_breaking_geometry}(b)].
The comparable rotations show that the valence-band Zeeman interaction is sufficient to generate the symmetry-selected response, while envelope-orbital magnetic coupling modifies its quantitative tensor elements.

Figure~\ref{fig:strain_yz_breaking_geometry}(c) shows that the canonical-vector angle reaches about \(78^\circ\) for the \(z\)-even shear profile, compared with maxima of about \(33^\circ\) and \(19^\circ\) for the \(\epsilon_{yy}\) and \(\epsilon_{zz}\) sweeps.
The broad large-\(\Theta_{LR}\) lobes over oblique field directions reflect the loss of a common set of principal axes between the two local \(g\) tensors.
The same differential rotation shapes the vector half mismatch in Fig.~\ref{fig:strain_yz_breaking_geometry}(d), which reaches about \(0.40\) for a field tilted between \(y\) and \(z\).
Under \(\epsilon_{yy}\) and \(\epsilon_{zz}\), the two tensors remain nearly coaxial and strain acts mainly through their principal values, so the largest mismatches remain near \(y\) and \(z\), respectively.
For the \(z\)-even shear profile, the noncoaxial tensor axes instead enhance the mismatch for oblique fields and shift its maximum into the \(y\)-\(z\) sector.

\begin{figure}[!t]
  \centering
  \includegraphics[width=\columnwidth]{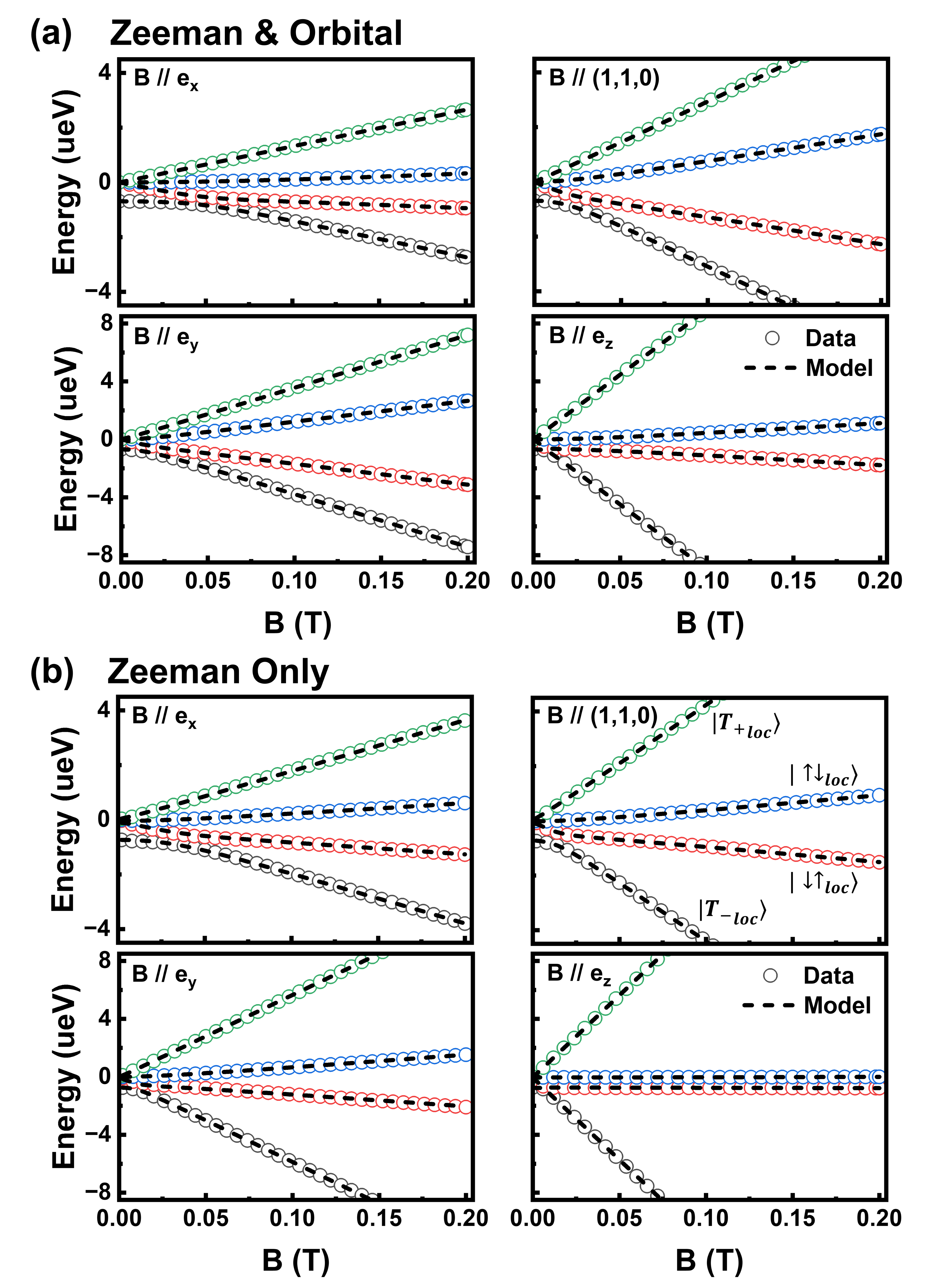}
  \caption{\label{fig:strained_spectrum}
  Directional two-hole spectra under the mirror-breaking, \(z\)-even \(\epsilon_{yz}=0.050\%\) profile.
  Panels (a) and (b) show the full and Zeeman-only Hamiltonians, respectively.
  The four field directions are the same as in the unstrained spectra of Fig.~\ref{fig:magnetic_spectrum}: \(\bm e_x\), \([110]\), \(\bm e_y\), and \(\bm e_z\).
  Open circles are CI eigenenergies and dashed curves are the fitted four-state-model branches.
  The state annotations in panel (b) indicate the approximate dominant local-spin configurations in the \(\Delta E_Z\)-dominated regime, away from avoided crossings.}
\end{figure}

Compared with the unstrained spectra in Fig.~\ref{fig:magnetic_spectrum}, the clearest change in Fig.~\ref{fig:strained_spectrum} is that the inner pair fans apart for most of the displayed field directions.
Figures~\ref{fig:strain_yz_breaking_components} and \ref{fig:strain_yz_breaking_geometry} identify the origin: the \(z\)-even shear opens the off-diagonal response and rotates the two local tensor frames relative to one another.
Their projected response magnitudes consequently differ over a broad range of field directions, so the scalar splitting mismatch \(\lvert\Delta E_Z\rvert\) grows rapidly with \(B\).
Once \(\lvert\Delta E_Z\rvert\) exceeds \(J_\perp\), the inner-branch separation
\(\Delta E_{\mathrm{inner}}\simeq\sqrt{J_\perp^2+\Delta E_Z^2}\)
is governed primarily by \(\lvert\Delta E_Z\rvert\).
The \(S_{\mathrm{loc}}\) and \(T_{0,\mathrm{loc}}\) labels used for the nearly parallel inner pair in the unstrained Zeeman-only spectrum are then replaced by the approximate local product states
\(\ket{\uparrow_L\downarrow_R}\) and \(\ket{\downarrow_L\uparrow_R}\).
Together with \(T_{+,\mathrm{loc}}\) and \(T_{-,\mathrm{loc}}\), these labels describe the four branches in Fig.~\ref{fig:strained_spectrum}(b) away from avoided crossings.

The full and Zeeman-only calculations both show this direction-dependent opening, most clearly for \(\bm B\parallel[110]\) and \(\bm B\parallel\bm e_y\), while the opening remains weaker for \(\bm B\parallel\bm e_z\).
Envelope-orbital magnetic coupling changes the absolute slopes and avoided-crossing details, but the Zeeman-only spectrum retains the shear-induced reorganization of the inner pair.

\begin{figure}[!t]
  \centering
  \includegraphics[width=\columnwidth]{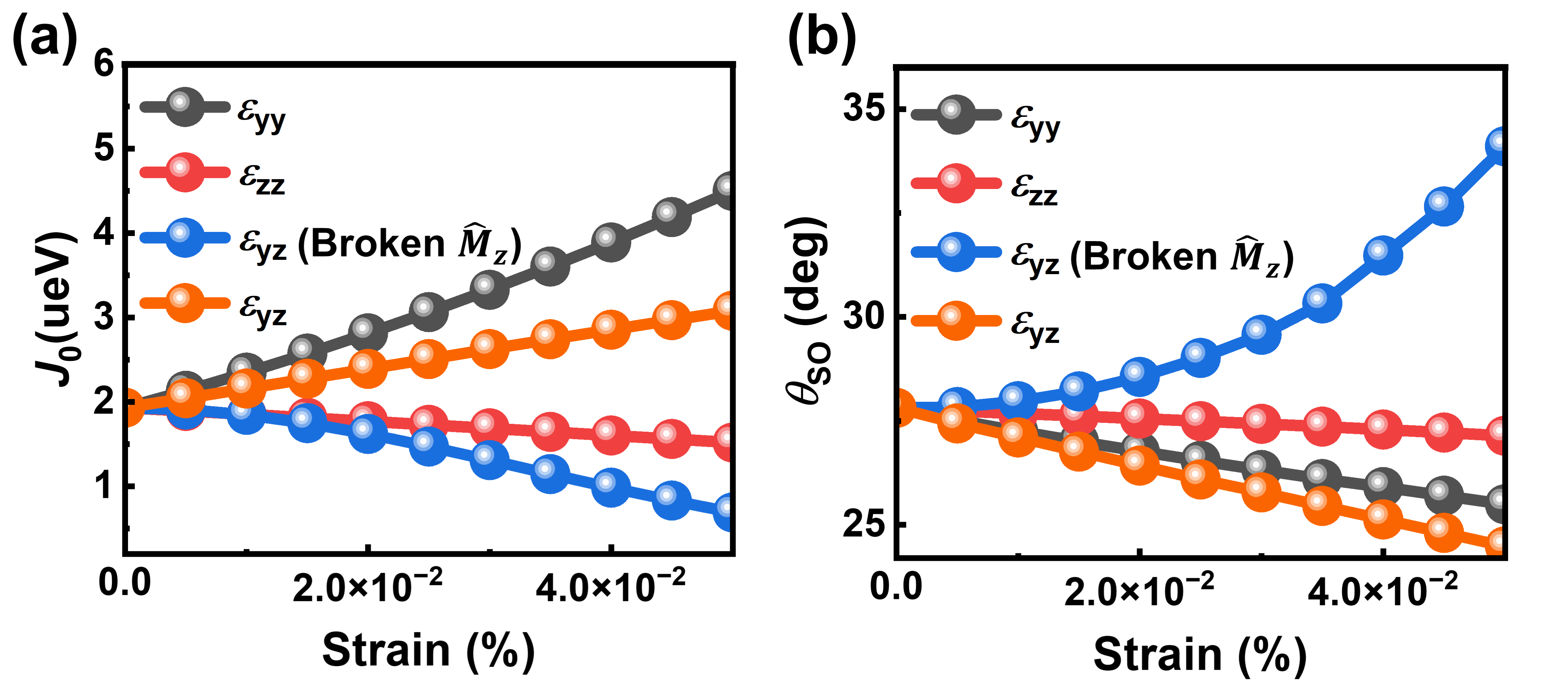}
  \caption{\label{fig:effective_parameters}
  Full-model strain dependence of the fitted exchange-sector parameters.
  Panel (a) shows the scale \(J_0\), and panel (b) the half-rotation angle \(\theta_{\mathrm{SO}}\), of the scaled-rotation exchange matrix \(\bm J=J_0\bm R(\bm n_{\mathrm{so}},-2\theta_{\mathrm{SO}})\).
  Gray, red, blue, and orange denote differential \(\epsilon_{yy}\), differential \(\epsilon_{zz}\), mirror-breaking \(z\)-even \(\epsilon_{yz}\), and mirror-compatible \(z\)-odd \(\epsilon_{yz}\), respectively.
  At each strain point, the magnetic-response tensors and the parameters of \(\bm J\) are fitted jointly to the same direction-resolved CI spectrum.}
\end{figure}

Figure~\ref{fig:effective_parameters} compares the fitted exchange sector across all four diagonal and shear strain protocols after the magnetic-response tensors have been fixed from the full-model spectrum fits.
Differential \(\epsilon_{yy}\) raises the fitted exchange scale \(J_0\), whereas \(\epsilon_{zz}\) lowers it more weakly.
For both diagonal protocols, \(\theta_{\mathrm{SO}}\) decreases more modestly.
The two shear profiles drive \(J_0\) and \(\theta_{\mathrm{SO}}\) in opposite directions: \(z\)-even shear lowers \(J_0\) and increases the half-rotation angle, while the \(z\)-odd profile does the reverse.
The largest energy RMSE among these joint fits is \(2.8\times10^{-3}~\ueV\).
These quantities parameterize the effective scaled-rotation exchange matrix over the fitted spectra.
Their microscopic identification with \(t_c\) and \(t_{\mathrm{so}}\) applies in the perturbative charge-elimination limit discussed in Appendix~\ref{app:effective}.

\section{Discussion}
\label{sec:discussion}

The results identify two distinct forms of magnetic-response control.
Differential \(\epsilon_{yy}\) and \(\epsilon_{zz}\) mainly change the principal \(g\) values while leaving the principal directions nearly unchanged, whereas the mirror-breaking \(z\)-even \(\epsilon_{yz}\) profile produces a large relative rotation of those directions.
This contrast follows from the valence-band coupling selected by the strain component together with the mirror constraint imposed by its spatial profile.
In the Pikus--Bir Hamiltonian, \(\epsilon_{yy}\) enters \(P_\epsilon\) and \(Q_\epsilon\), \(\epsilon_{zz}\) also activates \(R_\epsilon\), and \(\epsilon_{yz}\) enters \(S_\epsilon\).
The transverse parity of the profile determines whether \(M_z\) permits the off-diagonal elements that rotate the local magnetic response.

The relation between the tensor elements and these two types of response follows from the first-order strain correction to the local splitting tensor.
The strain amplitude is denoted by \(\lambda\), and \(\bm Q_i(\lambda)\) is the local splitting tensor of dot \(i\).
At zero strain, its principal directions \(\hat{\bm v}_{i,a}\) and squared principal values \(q_{i,a}=g_{i,a}^2\) satisfy
\(\bm Q_i(0)\hat{\bm v}_{i,a}=q_{i,a}\hat{\bm v}_{i,a}\).
In this principal basis,
\begin{equation}
\bm Q_i(\lambda)
=\operatorname{diag}(q_{i,1},q_{i,2},q_{i,3})
+\lambda\bm Q_i^{(1)}+O(\lambda^2),
\label{eq:q_linear_strain}
\end{equation}
where \(\bm Q_i^{(1)}=\left.\partial_\lambda\bm Q_i\right|_{\lambda=0}\).
When the three unstrained principal values \(q_{i,a}\) are distinct, first-order perturbation theory gives their strain-induced changes as
\begin{align}
\delta q_{i,a}
&=\lambda [\bm Q_i^{(1)}]_{aa},\nonumber\\
\delta\hat{\bm v}_{i,a}
&=\lambda\sum_{b\ne a}
\frac{[\bm Q_i^{(1)}]_{ba}}{q_{i,a}-q_{i,b}}
\hat{\bm v}_{i,b}.
\label{eq:q_axis_perturbation}
\end{align}
The diagonal entries of \(\bm Q_i^{(1)}\) therefore change the principal \(g\) values, while its off-diagonal entries rotate the principal directions.
For \(\epsilon_{yy}\) and \(\epsilon_{zz}\), this correction is predominantly diagonal, so strain changes the principal \(g\) values while the principal directions remain nearly unchanged.
Appendix~\ref{app:magnetoelastic} derives the same diagonal--off-diagonal structure from the Pikus--Bir and magnetic matrix elements.

The different behavior of the two shear profiles follows from the transformation of \(\bm Q_i(\lambda)\) under \(M_z\).
For the mirror-breaking \(z\)-even profile, \(M_z\) reverses the strain amplitude, \(\lambda\mapsto-\lambda\), and therefore
\begin{equation}
\bm Q_i(\lambda)=
\bm A_z^{\mathsf T}\bm Q_i(-\lambda)\bm A_z .
\label{eq:q_mirror_covariance}
\end{equation}
Here \(\bm A_z=\operatorname{diag}(-1,-1,1)\) represents the action of \(M_z\) on magnetic-field directions and reflects the axial-vector character of \(\bm B\).
The reflection does not exchange the dots, so Eq.~\eqref{eq:q_mirror_covariance} holds separately for each local tensor.
The first-order expansion of Eq.~\eqref{eq:q_mirror_covariance} restricts the strain correction to
\begin{equation}
\bm Q_i^{(1)}=
\begin{pmatrix}
0&0&c_{i,xz}\\
0&0&c_{i,yz}\\
c_{i,xz}&c_{i,yz}&0
\end{pmatrix}.
\label{eq:q_shear_linear_form}
\end{equation}
In the calculated tensor difference, the \(yz\) component is much larger than the \(xz\) component and produces the rotation in the \(y\)--\(z\) plane shown in Figs.~\ref{fig:strain_yz_breaking_components} and \ref{fig:strain_yz_breaking_geometry}.
For the mirror-compatible \(z\)-odd profile, the reflection instead leaves the strained system invariant at fixed \(\lambda\), giving
\begin{equation}
\bm Q_i(\lambda)=
\bm A_z^{\mathsf T}\bm Q_i(\lambda)\bm A_z,
\qquad
Q_{i,xz}=Q_{i,yz}=0 .
\label{eq:q_mirror_invariant}
\end{equation}
The \(xz\) and \(yz\) elements therefore vanish at each dot, excluding the first-order rotation channel and accounting for the nearly unchanged principal directions in Fig.~\ref{fig:strain_yz_preserving}.

Across all strain protocols, the Zeeman-only calculations reproduce the qualitative response found with the full Hamiltonian: \(\epsilon_{yy}\) and \(\epsilon_{zz}\) mainly reshape the principal \(g\) values, the mirror-compatible \(z\)-odd \(\epsilon_{yz}\) response remains weak, and the mirror-breaking \(z\)-even profile rotates the principal axes.
This agreement indicates that the strain-induced modification of the local \(g\) tensors arises primarily through the valence-band Zeeman term.
The comparison also shows that envelope-orbital magnetic coupling modifies the magnitude of the response and the detailed spectra while leaving these component- and symmetry-dependent trends unchanged.

\section{Conclusion}
\label{sec:conclusion}

We have investigated how differences in local strain reshape the magnetic response of a silicon FinFET double quantum dot with a three-dimensional Poisson--Schr\"odinger calculation based on a six-band \(k\!\cdot\!p\) model, configuration interaction, and a Fermi--Hubbard description of the correlated spectrum.
The diagonal components \(\epsilon_{yy}\) and \(\epsilon_{zz}\) act mainly on the principal values of the two local \(g\) tensors, with little opening of their maximum-response axes.
The response to \(\epsilon_{yz}\) is instead determined by the transverse mirror symmetry of its spatial profile.
The mirror-compatible \(z\)-odd profile suppresses the \(xz/yz\) rotation channel at each dot.
A mirror-breaking \(z\)-even profile produces a pronounced off-diagonal mismatch and rotates the principal axes of the two local responses relative to one another.
The resulting changes in local magnitude and orientation propagate to the inner two-hole branches of the correlated spectrum.

Their appearance in calculations containing only the valence-band Zeeman magnetic coupling shows that this interaction is sufficient to generate the trends, while the full Hamiltonian determines their magnitude and spectral expression.
Together, these results identify the strain component and transverse profile symmetry as complementary controls of whether coupled dots acquire a difference in the magnitude or orientation of their magnetic response.
Spatially resolved strain should therefore be included in realistic modeling of three-dimensional hole-spin devices and offers a route to engineer the magnetic anisotropy of coupled hole spins.

\begin{acknowledgments}
This work was supported by the National Natural Science Foundation of China under Grant Nos. T2293703 and T2293700 and by the 111 Project (B18001). We thank the High-Performance Computing Platform of Peking University and the National Supercomputer Center in Tianjin for providing computational resources.
\end{acknowledgments}

\appendix

\section{Six-band Luttinger--Kohn and Pikus--Bir matrix}
\label{app:lk}

The Luttinger--Kohn and Pikus--Bir operators are given in the six-band total-angular-momentum representation, which makes the orbital magnetic coupling and device-frame strain explicit.
The ordered basis comprises the heavy-hole, light-hole, and split-off states,
\begin{equation*}
\begin{aligned}
\mathcal B=\{&
\left|\tfrac32,\tfrac32\right\rangle,
\left|\tfrac32,\tfrac12\right\rangle,
\left|\tfrac32,-\tfrac12\right\rangle,\\
&
\left|\tfrac32,-\tfrac32\right\rangle,
\left|\tfrac12,\tfrac12\right\rangle,
\left|\tfrac12,-\tfrac12\right\rangle
\}.
\end{aligned}
\end{equation*}
In this basis, the multiband block used in the single-hole solver is
\begingroup
\setlength{\arraycolsep}{1.4pt}
\renewcommand{\arraystretch}{1.02}
\begin{equation}
\begin{gathered}
H_{\mathrm{LK+PB}}={}\\[-4pt]
\begin{pmatrix}
P+Q & -S & R & 0 & -S/\sqrt2 & \sqrt2 R \\
-S^{*} & P-Q & 0 & R & -\sqrt2 Q & \sqrt{3/2}S \\
R^{*} & 0 & P-Q & S & \sqrt{3/2}S^{*} & \sqrt2 Q \\
0 & R^{*} & S^{*} & P+Q & -\sqrt2 R^{*} & -S^{*}/\sqrt2 \\
-S^{*}/\sqrt2 & -\sqrt2 Q & \sqrt{3/2}S & -\sqrt2 R & P+\Delta_{\mathrm{so}} & 0 \\
\sqrt2 R^{*} & \sqrt{3/2}S^{*} & \sqrt2 Q & -S/\sqrt2 & 0 & P+\Delta_{\mathrm{so}}
\end{pmatrix}
\end{gathered},
\label{eq:lk_full}
\end{equation}
\endgroup
where \(P=P_k+P_\epsilon\), \(Q=Q_k+Q_\epsilon\),
\(R=R_k+R_\epsilon\), and \(S=S_k+S_\epsilon\).
In terms of the crystal-frame covariant components \(\pi_X,\pi_Y,\pi_Z\), the kinetic terms are
\begin{align}
P_k&=\frac{\hbar^2}{2m_0}\gamma_1
\left(\pi_X^2+\pi_Y^2+\pi_Z^2\right),\\
Q_k&=\frac{\hbar^2}{2m_0}\gamma_2
\left(\pi_X^2+\pi_Y^2-2\pi_Z^2\right),\\
R_k&=\frac{\sqrt3\hbar^2}{2m_0}
\left[
-\gamma_2(\pi_X^2-\pi_Y^2)
+2i\gamma_3\pi_X\pi_Y
\right],\\
S_k&=\frac{\sqrt3\hbar^2}{m_0}\gamma_3
(\pi_X-i\pi_Y)\pi_Z .
\end{align}
The crystal and device covariant components satisfy
\(\bm\pi_{\mathrm{crys}}=\bm R_{cd}\bm\pi_{\mathrm{dev}}\).
The vector potential is specified in device coordinates.
With \(e>0\), the covariant derivative is
\(\bm\pi_{\mathrm{dev}}=-i\bm\nabla_{\mathrm{dev}}+(e/\hbar)\bm A\).
For a uniform field \(\bm B=(B_x,B_y,B_z)\), the selected gauge is
\begin{equation}
\bm A
=\left(-\frac{B_z y}{2},\frac{B_z x}{2},B_x y-B_y x\right),
\qquad
\bm\nabla\times\bm A=\bm B .
\label{eq:uniform_field_gauge}
\end{equation}
which combines the symmetric gauge for \(B_z\) with an \(A_z\) representation of the in-plane field.
Because the covariant derivatives do not commute at finite magnetic field, terms such as \(\pi_X\pi_Y\) in \(R_k\) and \((\pi_X-i\pi_Y)\pi_Z\) in \(S_k\) require an ordering prescription.
They are implemented in Hermitian symmetrized form, for example \(\pi_X\pi_Y\mapsto\{\pi_X,\pi_Y\}/2\), through the finite-element weak formulation.
The Pikus--Bir combinations in Eq.~\eqref{eq:lk_full} are likewise written in the cubic crystal frame:
\begin{align}
P_\epsilon&=a_v(\epsilon_{XX}+\epsilon_{YY}+\epsilon_{ZZ}),\\
Q_\epsilon&=b\left[
\epsilon_{ZZ}-\frac{\epsilon_{XX}+\epsilon_{YY}}{2}
\right],\\
R_\epsilon&=-\frac{\sqrt3}{2}b
(\epsilon_{XX}-\epsilon_{YY})-i d\epsilon_{XY},\\
S_\epsilon&=-d(\epsilon_{XZ}-i\epsilon_{YZ}).
\label{eq:pikus_bir_terms}
\end{align}
To evaluate these couplings for the device-frame strain profiles, we first rotate the strain tensor into the cubic crystal frame using Eq.~\eqref{eq:device_crystal_rotation}.
With one nonzero device-frame strain component, the resulting Pikus--Bir channels are
\begin{equation}
\begin{aligned}
\epsilon_{yy}^{\mathrm{dev}}=\lambda:\quad
&(P_\epsilon,Q_\epsilon,R_\epsilon,S_\epsilon)
=(a_v\lambda,b\lambda,0,0),\\
\epsilon_{zz}^{\mathrm{dev}}=\lambda:\quad
&(P_\epsilon,Q_\epsilon,R_\epsilon,S_\epsilon)
=(a_v\lambda,-b\lambda/2,id\lambda/2,0),\\
\epsilon_{yz}^{\mathrm{dev}}=\lambda:\quad
&(P_\epsilon,Q_\epsilon,R_\epsilon,S_\epsilon)
=\left(0,0,0,-\frac{d(1+i)}{\sqrt2}\lambda\right).
\end{aligned}
\label{eq:device_pb_channels}
\end{equation}
Thus, \(\epsilon_{yy}^{\mathrm{dev}}\) couples through \(P_\epsilon\) and \(Q_\epsilon\), \(\epsilon_{zz}^{\mathrm{dev}}\) additionally generates the complex \(R_\epsilon\) channel, and \(\epsilon_{yz}^{\mathrm{dev}}\) couples exclusively through \(S_\epsilon\)~\cite{BirPikus1974,Winkler2003,Wang2024}.
\section{Linear magnetoelastic response of a Kramers doublet}
\label{app:magnetoelastic}

The derivation starts from the zero-strain, zero-field Hamiltonian \(H_0\) of a selected localized one-hole state.
The projector \(\mathcal P\) selects its lowest Kramers doublet with energy \(E_0\), and the reduced resolvent is
\begin{equation}
\mathcal R=(1-\mathcal P)
\frac{1}{E_0-H_0}
(1-\mathcal P).
\label{eq:reduced_resolvent}
\end{equation}
For a strain amplitude \(\lambda\), the corresponding perturbation operator is
\(D=\left.\partial H/\partial\lambda\right|_{\lambda=\bm B=0}\), and the linear magnetic-response operator is decomposed as
\begin{equation}
\mathcal M_j=
\left.\frac{\partial H}{\partial B_j}\right|_{\lambda=\bm B=0}
=\mathcal M_j^{\mathrm{BZ}}+\mathcal M_j^{\mathrm{orb}},
\label{eq:magnetic_operator_channels}
\end{equation}
where the two terms originate from \(H_Z\) and from the vector potential in \(H_{\mathrm{LK}}\), respectively.
Time-reversal symmetry keeps the isolated Kramers doublet degenerate at \(B=0\), giving
\(\mathcal P D\mathcal P=d_0\mathcal P\).
Generalizing the mixed electric--magnetic response structure of Ref.~\cite{Venitucci2018} to the strain perturbation \(D\), L\"owdin partitioning gives the leading strain-dependent spin term at order \(O(\lambda B)\):
\begin{equation}
\begin{aligned}
H_{\mathrm{eff}}^{(\lambda B)}
=\lambda\sum_jB_j\,\mathcal P
\bigl[
&D\mathcal R\mathcal M_j
+\mathcal M_j\mathcal R D\\
&+\partial_\lambda\mathcal M_j
\bigr]\mathcal P .
\end{aligned}
\label{eq:mixed_strain_field_response}
\end{equation}
In the model used here, strain enters through the Pikus--Bir potential with fixed material parameters, so \(\partial_\lambda\mathcal M_j=0\).

In a parallel-transported Kramers basis, \(\widetilde{\bm G}\) is the microscopic projected \(g\)-matrix.
For either magnetic channel \(x\in\{\mathrm{BZ},\mathrm{orb}\}\),
\begin{equation}
\begin{aligned}
\frac{\partial \widetilde G_{aj}^{x}}{\partial\lambda}
=\frac{1}{\mub}\Tr\biggl\{
\sigma_a\mathcal P
\bigl[
&D\mathcal R\mathcal M_j^{x}
+\mathcal M_j^{x}\mathcal R D\\
&+\partial_\lambda\mathcal M_j^{x}
\bigr]\mathcal P
\biggr\}.
\end{aligned}
\label{eq:microscopic_g_derivative}
\end{equation}
The matrix \(\widetilde{\bm G}\) is real because the projected Zeeman Hamiltonian is Hermitian.
Its rows transform with the Kramers basis, while
\(\partial_\lambda(\widetilde{\bm G}^{\mathsf T}\widetilde{\bm G})=\partial_\lambda\bm Q\) is basis invariant.
Equation~\eqref{eq:microscopic_g_derivative} pairs each Pikus--Bir matrix element with the Bloch-Zeeman and envelope-orbital magnetic matrix elements.
These channel-resolved derivatives refer to the microscopic matrix \(\widetilde{\bm G}\).
The figures use the canonical matrices \(\bm G_i=\bm Q_i^{1/2}\) constructed from the total splitting tensors.
In the common Kramers basis, \(\widetilde{\bm G}=\widetilde{\bm G}^{\mathrm{BZ}}+\widetilde{\bm G}^{\mathrm{orb}}\), and therefore
\begin{equation}
\begin{aligned}
\bm Q
={}&
\bigl(\widetilde{\bm G}^{\mathrm{BZ}}\bigr)^{\mathsf T}
\widetilde{\bm G}^{\mathrm{BZ}}
+
\bigl(\widetilde{\bm G}^{\mathrm{orb}}\bigr)^{\mathsf T}
\widetilde{\bm G}^{\mathrm{orb}}\\
&+
\bigl(\widetilde{\bm G}^{\mathrm{BZ}}\bigr)^{\mathsf T}
\widetilde{\bm G}^{\mathrm{orb}}
+
\bigl(\widetilde{\bm G}^{\mathrm{orb}}\bigr)^{\mathsf T}
\widetilde{\bm G}^{\mathrm{BZ}} .
\end{aligned}
\label{eq:q_channel_interference}
\end{equation}
The full response is set by the two channel self terms and the two Bloch--orbital interference terms in Eq.~\eqref{eq:q_channel_interference}.
The Zeeman-only calculation evaluates the structure generated by the Bloch-Zeeman part.
Subtracting the Zeeman-only splitting tensor from the full splitting tensor leaves the orbital self term and the two Bloch--orbital cross terms in Eq.~\eqref{eq:q_channel_interference}.
The corresponding difference between their canonical square roots is a nonlinear full/Zeeman-only comparison rather than an additive channel contribution.
Equation~\eqref{eq:microscopic_g_derivative} applies to a selected localized one-hole doublet, while the tensors in the main text are the corresponding low-energy parameters extracted from the correlated double-dot spectrum.

Equations~\eqref{eq:mixed_strain_field_response} and \eqref{eq:microscopic_g_derivative} apply to all three strain-tensor components and, for \(\epsilon_{yz}\), to either spatial profile.
For \(u\in\{yy,zz,yz\}\), the first-order microscopic response is
\(\widetilde{\bm G}^{(1)}_u=
\left.\partial_{\lambda_u}\widetilde{\bm G}\right|_0\),
with \(D_u\) formed by combining the corresponding Pikus--Bir channel in Eq.~\eqref{eq:device_pb_channels} with the spatial profile defined in Sec.~\ref{sec:device_method}.
The unstrained microscopic reference in the same parallel-transported Kramers basis is
\(\widetilde{\bm G}_{\mathrm{ref}}\equiv
\left.\widetilde{\bm G}\right|_{\lambda=0}\).
The canonical tensor used in the figures is \(\bm G_i=\bm Q_i^{1/2}\).
The corresponding first-order change of the splitting tensor is
\begin{equation}
\bm Q_u^{(1)}
=
\bigl(\widetilde{\bm G}^{(1)}_u\bigr)^{\mathsf T}
\widetilde{\bm G}_{\mathrm{ref}}
+
\widetilde{\bm G}_{\mathrm{ref}}^{\mathsf T}
\widetilde{\bm G}^{(1)}_u .
\label{eq:microscopic_q_derivative}
\end{equation}
When \(\bm Q_u^{(1)}\) is expressed in the eigenbasis of the unstrained tensor and its principal values are distinct, the diagonal entries generate the first-order principal-value shifts, whereas the off-diagonal entries generate the first-order axis rotations described by Eq.~\eqref{eq:q_axis_perturbation}.

For device-frame \(\epsilon_{yy}=\lambda_{yy}\), the perturbation contains \(P_\epsilon=a_v\lambda_{yy}\) and \(Q_\epsilon=b\lambda_{yy}\), with \(R_\epsilon=S_\epsilon=0\).
The localized \(P_\epsilon\) term shifts the confinement landscape.
Its spatially uniform part gives a common band-edge shift within an isolated dot.
Within the \(J=3/2\) subblock, \(Q_\epsilon\) changes the heavy-hole--light-hole separation, and the complete six-band matrix also includes the associated split-off admixture.
Their projected first-order effect is contained in the \(D_{yy}\) matrix elements and the resolvent \(\mathcal R\) of Eq.~\eqref{eq:mixed_strain_field_response}.
For the confined states calculated here, this projection is dominated by diagonal entries of \(\bm Q_{yy}^{(1)}\), producing the principal-value redistribution seen in Fig.~\ref{fig:strain_yy}.

For device-frame \(\epsilon_{zz}=\lambda_{zz}\), one instead has
\(P_\epsilon=a_v\lambda_{zz}\),
\(Q_\epsilon=-b\lambda_{zz}/2\), and
\(R_\epsilon=i d\lambda_{zz}/2\).
The additional cubic valence-band matrix element \(R_\epsilon\) changes the heavy-hole--light-hole/split-off admixture before projection into \(\bm Q\).
The resulting \(\bm Q_{zz}^{(1)}\) is again dominated by principal-value changes, but with a component ordering and magnitude different from the \(\epsilon_{yy}\) response in Fig.~\ref{fig:strain_zz}.
Both magnetic operators \(\mathcal M_j^{\mathrm{BZ}}\) and \(\mathcal M_j^{\mathrm{orb}}\) enter the symmetry-allowed mixed paths in Eq.~\eqref{eq:mixed_strain_field_response}.
The Zeeman-only response follows from \(\mathcal M_j^{\mathrm{orb}}=0\) and has the same predominantly diagonal structure with different coefficients.

For device-frame \(\epsilon_{yz}=\lambda\), Eq.~\eqref{eq:device_pb_channels} gives an \(S_\epsilon\) term that couples heavy-hole and light-hole or split-off components.
When the local \(M_z\) constraint is broken, components with the same lowest envelope symmetry are no longer forbidden from having a finite overlap.
One symmetry-allowed Bloch-Zeeman contribution within this envelope sector, labeled \(0\), is the virtual path
\begin{equation}
\lvert\mathrm{HH},0\rangle
\xrightarrow{\ D\ }
\lvert\mathrm{LH/SO},0\rangle
\xrightarrow{\ \mathcal M_z^{\mathrm{BZ}}\ }
\lvert\mathrm{HH},0\rangle .
\label{eq:bloch_shear_path}
\end{equation}
This path closes without an additional envelope excitation.
In a minimal heavy-hole--light-hole reduction, \(\bm\tau\) acts in the projected Kramers doublet.
The mixed strain--field term generated by this path has the form
\begin{equation}
H_{\mathrm{HH,eff}}
\supset
\frac{\mub}{2}
\left(g_yB_y+v\lambda B_z\right)\tau_z ,
\label{eq:minimal_shear_response}
\end{equation}
where \(v\) is set by deformation-potential matrix elements and valence-band energy denominators.
The pre-existing \(B_y\) response and the strain-induced \(B_z\) response are collinear in this pseudospin space.
The corresponding \(y\)--\(z\) field-space block of the splitting tensor is
\begin{equation}
\bm Q_{yz}^{\mathrm{min}}=
\begin{pmatrix}
g_y^2&g_yv\lambda\\
g_yv\lambda&v^2\lambda^2
\end{pmatrix}.
\label{eq:minimal_shear_q}
\end{equation}
Hence \(Q_{yz}=O(\lambda)\), the axis rotation begins at \(O(\lambda)\), and the principal-value shift produced by this off-diagonal path begins at \(O(\lambda^2)\).
For the \(z\)-odd profile, the local mirror parity enforces the invariant form in Eq.~\eqref{eq:q_mirror_invariant} and removes this \(Q_{yz}\) matrix element in the ideal mirror-symmetric limit.
The coefficients of the complete six-band Hamiltonian and the corresponding energy denominators set the magnitude of the allowed path.
Envelope-orbital magnetic coupling is known to renormalize the hole magnetic response through valence-band mixing~\cite{Ares2013}.
In the present decomposition, this coupling appears through the orbital self term and the Bloch--orbital interference terms in Eq.~\eqref{eq:q_channel_interference}.

\section{CI matrix elements and state labels}
\label{app:ci}

For orthonormal single-hole spinors, Eq.~\eqref{eq:slater} defines an orthonormal determinant basis indexed by \(I=(m,n)\) with \(m<n\).
The matrix elements can be written as
\begin{equation}
\left(H_{\mathrm{CI}}\right)_{IJ}
=
(\varepsilon_m+\varepsilon_n)\delta_{IJ}
+\left\langle mn\middle|V_C\middle|pq\right\rangle
-\left\langle mn\middle|V_C\middle|qp\right\rangle ,
\label{eq:ci_matrix_elements}
\end{equation}
where \(J=(p,q)\).
Charge labels in the main text refer to the dominant \((2,0)\), \((1,1)\), or \((0,2)\) occupation of a CI eigenstate.
Spin--orbit coupling removes exact total-spin quantum numbers, so ``singlet-like'' and ``triplet-like'' denote the dominant character and its adiabatic continuation through hybridized regions.
The states \(\ket S\) and \(\ket{T_m}\) introduced below are basis states of the reduced two-site pseudospin model.

\begin{widetext}
\section{Five-state Fermi--Hubbard reduction}
\label{app:effective}

Equation~\eqref{eq:two_site_model} separates into charge, spin-conserving tunneling, spin--orbit tunneling, and Zeeman contributions:
\begin{align}
H_{\mathrm{FH}}={}&H_{\mathrm{ch}}+H_t+H_{\mathrm{so}}+H_Z,
\label{eq:fh_components}\\
H_{\mathrm{ch}}={}&\sum_{i=L,R}\left(\varepsilon_i n_i+U_i n_{i\uparrow}n_{i\downarrow}\right)+U_{LR}n_Ln_R,
\label{eq:fh_charge}\\
H_t={}&t_c\sum_s\left(c_{Ls}^{\dagger}c_{Rs}+c_{Rs}^{\dagger}c_{Ls}\right),
\label{eq:fh_tunnel}\\
H_{\mathrm{so}}={}&-i t_{\mathrm{so}}\sum_{ss'}c_{Ls}^{\dagger}
\left(\bm n_{\mathrm{so}}\cdot\bm\sigma\right)_{ss'}c_{Rs'}+\mathrm{H.c.},
\label{eq:fh_so}\\
H_Z={}&\frac12\sum_{i=L,R}\sum_{ss'}c_{is}^{\dagger}
\left(\bm\sigma\cdot\bm b_i\right)_{ss'}c_{is'} .
\label{eq:fh_zeeman}
\end{align}
Here \(n_i=\sum_s c_{is}^{\dagger}c_{is}\) and \(\bm b_i=\mub B\,\bm G_i\hat{\bm n}\) is the linear Zeeman field, as in the main text.
The displayed charge ordering takes the retained doubly occupied configuration to be \(\ket{S_d}=\ket{S(0,2)}\).
Exchanging \(L\) and \(R\) gives the equivalent form on the opposite side of the charge anticrossing.
The five states used for the projection are
\begin{equation}
\begin{aligned}
\ket{S_d}&=c_{R\uparrow}^{\dagger}c_{R\downarrow}^{\dagger}\ket{0},\\
\ket{S}&=\frac{c_{L\uparrow}^{\dagger}c_{R\downarrow}^{\dagger}-c_{L\downarrow}^{\dagger}c_{R\uparrow}^{\dagger}}{\sqrt2}\ket{0},\\
\ket{T_0}&=\frac{c_{L\uparrow}^{\dagger}c_{R\downarrow}^{\dagger}+c_{L\downarrow}^{\dagger}c_{R\uparrow}^{\dagger}}{\sqrt2}\ket{0},\\
\ket{T_+}&=c_{L\uparrow}^{\dagger}c_{R\uparrow}^{\dagger}\ket{0},
\qquad
\ket{T_-}=c_{L\downarrow}^{\dagger}c_{R\downarrow}^{\dagger}\ket{0}.
\end{aligned}
\label{eq:five_state_basis}
\end{equation}
Thus \(\mathcal B_5=\{\ket{S_d}\}\cup\mathcal B_4\), where
\(\mathcal B_4=\{\ket{S},\ket{T_0},\ket{T_+},\ket{T_-}\}\).
With the separated-charge singlet as the charge-sector reference, the retained double-occupation energy is
\begin{equation}
\Delta_d=E(S_d)-E(S)=\varepsilon_R-\varepsilon_L+U_R-U_{LR}.
\label{eq:double_occupation_energy}
\end{equation}
The same expression with \(L\) and \(R\) exchanged applies when \(\ket{S(2,0)}\) is retained.

The mean and difference fields are
\(\bar{\bm b}=(\bm b_L+\bm b_R)/2\) and
\(\delta\bm b=(\bm b_L-\bm b_R)/2\), respectively.
For any of the vectors
\(\bm a\in\{\bar{\bm b},\delta\bm b,\bm n_{\mathrm{so}}\}\), the notation \(a_{\pm}=a_x\pm i a_y\) is used.
Direct evaluation of Eqs.~\eqref{eq:fh_tunnel}--\eqref{eq:fh_zeeman} in \(\mathcal B_5\) gives
\begin{equation}
H_5=
\begin{pmatrix}
\Delta_d & \sqrt2\,t_c & +i\sqrt2\,t_{\mathrm{so}}n_{\mathrm{so},z} & -i t_{\mathrm{so}}n_{\mathrm{so},+} & +i t_{\mathrm{so}}n_{\mathrm{so},-}\\[2pt]
\sqrt2\,t_c & 0 & \delta b_z & -\delta b_+/\sqrt2 & +\delta b_-/\sqrt2\\[2pt]
-i\sqrt2\,t_{\mathrm{so}}n_{\mathrm{so},z} & \delta b_z & 0 & \bar b_+/\sqrt2 & \bar b_-/\sqrt2\\[2pt]
+i t_{\mathrm{so}}n_{\mathrm{so},-} & -\delta b_-/\sqrt2 & \bar b_-/\sqrt2 & \bar b_z & 0\\[2pt]
-i t_{\mathrm{so}}n_{\mathrm{so},+} & +\delta b_+/\sqrt2 & \bar b_+/\sqrt2 & 0 & -\bar b_z
\end{pmatrix}.
\label{eq:five_state_matrix}
\end{equation}
The first row and column isolate the charge-excited state, so Eq.~\eqref{eq:five_state_matrix} can equivalently be written as
\begin{equation}
H_5=
\begin{pmatrix}
\Delta_d & \bm v^\dagger\\
\bm v & H_Z^{(11)}
\end{pmatrix},
\qquad
\bm v=
\begin{pmatrix}
\sqrt2\,t_c\\
-i\sqrt2\,t_{\mathrm{so}}n_{\mathrm{so},z}\\
+i\,t_{\mathrm{so}}n_{\mathrm{so},-}\\
-i\,t_{\mathrm{so}}n_{\mathrm{so},+}
\end{pmatrix}.
\label{eq:five_state_block}
\end{equation}
The four-state block appearing here is
\begin{equation}
H_Z^{(11)}=
\begin{pmatrix}
0 & \delta b_z & -\delta b_+/\sqrt2 & +\delta b_-/\sqrt2\\
\delta b_z & 0 & \bar b_+/\sqrt2 & \bar b_-/\sqrt2\\
-\delta b_-/\sqrt2 & \bar b_-/\sqrt2 & \bar b_z & 0\\
+\delta b_+/\sqrt2 & \bar b_+/\sqrt2 & 0 & -\bar b_z
\end{pmatrix}.
\label{eq:four_state_zeeman}
\end{equation}
The mean local response acts within the triplet sector, while the local-response difference couples the separated-charge singlet to the triplets.
At \(B=0\), \(\bm v^\dagger\bm v=2(t_c^2+t_{\mathrm{so}}^2)=2t^2\).
Within this five-state truncation, the exact separation between the three uncoupled zero-energy combinations and the hybridized ground state is
\begin{equation}
J_{\mathrm{gap}}
=\frac{\sqrt{\Delta_d^2+8(t_c^2+t_{\mathrm{so}}^2)}-\Delta_d}{2}.
\label{eq:five_state_zero_field_gap}
\end{equation}

An explicit intermediate low-energy basis follows by diagonalizing the charge-singlet subblock at \(B=t_{\mathrm{so}}=0\):
\begin{equation}
H_{\mathrm{ch}}^{(S)}=
\begin{pmatrix}
\Delta_d & \sqrt2\,t_c\\
\sqrt2\,t_c & 0
\end{pmatrix},
\qquad
E_{\pm}=\frac{\Delta_d\pm\sqrt{\Delta_d^2+8t_c^2}}{2}.
\label{eq:charge_singlet_block}
\end{equation}
For \(\Delta_d>0\), the mixing angle is specified by \(\tan\gamma=2\sqrt2t_c/\Delta_d\), with \(c_\gamma=\cos(\gamma/2)\) and \(s_\gamma=\sin(\gamma/2)\).
The charge eigenstates can then be chosen as \(\ket{S_-}=c_\gamma\ket{S}-s_\gamma\ket{S_d}\) and \(\ket{S_+}=s_\gamma\ket{S}+c_\gamma\ket{S_d}\), with the spin-conserving charge-singlet offset \(J_{\mathrm{c}}=-E_-=[\sqrt{\Delta_d^2+8t_c^2}-\Delta_d]/2\).
When the upper charge state \(\ket{S_+}\) is well separated from the spin manifold, projecting it out and retaining \(\ket{S_-}\) with the three triplets gives
\begin{equation}
H_{\mathrm{4,mix}}=
\begin{pmatrix}
-J_{\mathrm{c}} & \beta_0 & \beta_+ & \beta_-\\
\beta_0^* & 0 & \bar b_+/\sqrt2 & \bar b_-/\sqrt2\\
\beta_+^* & \bar b_-/\sqrt2 & \bar b_z & 0\\
\beta_-^* & \bar b_+/\sqrt2 & 0 & -\bar b_z
\end{pmatrix},
\label{eq:mixed_four_state}
\end{equation}
where
\begin{equation}
\begin{aligned}
\beta_0&=c_\gamma\,\delta b_z-i\sqrt2s_\gamma t_{\mathrm{so}}n_{\mathrm{so},z},\\
\beta_+&=-\frac{c_\gamma}{\sqrt2}\delta b_++i s_\gamma t_{\mathrm{so}}n_{\mathrm{so},+},\\
\beta_-&=+\frac{c_\gamma}{\sqrt2}\delta b_--i s_\gamma t_{\mathrm{so}}n_{\mathrm{so},-}.
\end{aligned}
\label{eq:mixed_couplings}
\end{equation}
Equation~\eqref{eq:mixed_four_state} displays the two physical paths into the singlet--triplet couplings: the differential Zeeman response carries the separated-charge weight \(c_\gamma\), whereas spin-flip tunneling carries the double-occupation weight \(s_\gamma\).

The same virtual processes can be represented directly in the separated-charge basis.
A five-state eigenvector has the form \(\ket{\Psi}=q\ket{S_d}+\ket{\psi_4}\), where \(\ket{\psi_4}=P_{11}\ket{\Psi}\) and \(P_{11}=\sum_{\alpha\in\mathcal B_4}\ket{\alpha}\bra{\alpha}\).
The \(\ket{S_d}\) component of \(H_5\ket{\Psi}=E\ket{\Psi}\) gives \(q=(E-\Delta_d)^{-1}\bm v^\dagger\ket{\psi_4}\).
Substitution into the four-state component gives the exact energy-dependent L\"owdin--Feshbach reduction
\begin{align}
H_{\mathrm{4s}}(E)
={}&H_Z^{(11)}+\frac{\bm v\bm v^\dagger}{E-\Delta_d}\nonumber\\
={}&H_Z^{(11)}-\frac{\bm v\bm v^\dagger}{\Delta_d-E}.
\label{eq:sw_projection}
\end{align}
For reference, the second-order charge-mediated contribution is the explicit matrix
\begin{equation}
H_{\mathrm{ex}}(E)=-\frac{\mathcal M}{\Delta_d-E},
\qquad
\mathcal M=\bm v\bm v^\dagger=
\begin{pmatrix}
2t_c^2 & +2i t_c t_{\mathrm{so}}n_{\mathrm{so},z} & -i\sqrt2\,t_c t_{\mathrm{so}}n_{\mathrm{so},+} & +i\sqrt2\,t_c t_{\mathrm{so}}n_{\mathrm{so},-}\\[2pt]
-2i t_c t_{\mathrm{so}}n_{\mathrm{so},z} & 2t_{\mathrm{so}}^2n_{\mathrm{so},z}^2 & -\sqrt2\,t_{\mathrm{so}}^2n_{\mathrm{so},z}n_{\mathrm{so},+} & +\sqrt2\,t_{\mathrm{so}}^2n_{\mathrm{so},z}n_{\mathrm{so},-}\\[2pt]
+i\sqrt2\,t_c t_{\mathrm{so}}n_{\mathrm{so},-} & -\sqrt2\,t_{\mathrm{so}}^2n_{\mathrm{so},z}n_{\mathrm{so},-} & t_{\mathrm{so}}^2n_{\mathrm{so},-}n_{\mathrm{so},+} & -t_{\mathrm{so}}^2n_{\mathrm{so},-}^2\\[2pt]
-i\sqrt2\,t_c t_{\mathrm{so}}n_{\mathrm{so},+} & +\sqrt2\,t_{\mathrm{so}}^2n_{\mathrm{so},z}n_{\mathrm{so},+} & -t_{\mathrm{so}}^2n_{\mathrm{so},+}^2 & t_{\mathrm{so}}^2n_{\mathrm{so},+}n_{\mathrm{so},-}
\end{pmatrix}.
\label{eq:exchange_matrix}
\end{equation}
With the tunneling matrix of Eq.~\eqref{eq:tunneling_angle}, the second-order exchange block in the separated-charge product basis takes the rotated form of Eq.~\eqref{eq:exchange_rotation}.
For \(|E|,t\ll\Delta_d\), Eq.~\eqref{eq:sw_projection} reduces to the usual Schrieffer--Wolff limit~\cite{SchriefferWolff1966},
\begin{equation}
J_0\simeq\frac{2(t_c^2+t_{\mathrm{so}}^2)}{\Delta_d},
\qquad
\bm J=J_0\bm R(\bm n_{\mathrm{so}},-2\theta_{\mathrm{SO}}).
\label{eq:hubbard_exchange_rotation}
\end{equation}
Within this Schrieffer--Wolff mapping, the lab-frame exchange tensor carries the spin-space rotation \(-2\theta_{\mathrm{SO}}\).
\(J_{\mathrm{c}}\) equals \(J_{\mathrm{gap}}\) when \(t_{\mathrm{so}}=0\), whereas \(J_0\simeq2t^2/\Delta_d\) approaches \(J_{\mathrm{gap}}\) in the perturbative limit \(t/\Delta_d\ll1\).
For \(t_{\mathrm{so}}=0\) and Zeeman energies small compared with \(\Delta_d\), this term lowers only the separated-charge singlet.  With \(\bm S_i=\bm\sigma_i/2\), it becomes
\begin{align}
H_{\mathrm{ex}}&\simeq-J_{\mathrm{FH}}\ket{S}\bra{S}
=J_{\mathrm{FH}}\left(\bm S_L\cdot\bm S_R-\frac14\right),\nonumber\\
J_{\mathrm{FH}}&\simeq\frac{2t_c^2}{\Delta_d}.
\label{eq:exchange_limit}
\end{align}
Thus \(J_{\mathrm{FH}}\) is the spin-conserving limit of \(J_0\) at this order.
At finite \(t_{\mathrm{so}}\), the off-diagonal entries of \(\mathcal M\) generate spin--orbit-assisted singlet--triplet couplings.
Equations~\eqref{eq:mixed_couplings} and \eqref{eq:sw_projection} express the two low-energy mixing paths: the differential local Zeeman field acts within the separated-charge sector, while spin-flip tunneling enters through virtual double occupation.
\end{widetext}

\section{Spectral sampling and exchange fits}
\label{app:spectral_sampling}

The full-model strain series share an unstrained reference and contain ten nonzero points from \(0.005\%\) to \(0.050\%\).
The matched full/Zeeman-only shear comparison uses the ten common nonzero points.
The unstrained tensor fit samples 33 magnetic-field directions over \(0\)--\(0.27~\mathrm T\).
The strained fits likewise use 33 directions, with component-dependent field windows extending from \(0.18~\mathrm T\) to \(0.30\)--\(0.70~\mathrm T\).
Each field direction carries the same total fitting weight.

For Fig.~\ref{fig:effective_parameters}, the local magnetic-response tensors and the scaled-rotation exchange matrix are fitted jointly to the direction-resolved CI eigenenergies at each strain point.
The reported exchange parameters are therefore effective fit parameters; their microscopic relation to tunneling amplitudes applies in the charge-elimination limit derived in Appendix~\ref{app:effective}.

\section{Extraction and canonical representation of the local magnetic response}
\label{app:q_fit}

In the direction-resolved spectral fit, the local response of each site is parameterized by
\begin{equation}
g_i^2(\hat{\bm n}_a)
=\hat{\bm n}_a^{\mathsf T}
\bm Q_i
\hat{\bm n}_a ,
\label{eq:q_fit}
\end{equation}
with a positive-definite \(\bm Q_i\).
The eigenvalues and eigenvectors of \(\bm Q_i\) give its principal response values and field directions.

The microscopic transformation
\(\widetilde{\bm G}_i\rightarrow\bm O_i\widetilde{\bm G}_i\), with
\(\bm O_i\in SO(3)\) the proper spin-space rotation induced by a Kramers-basis change, leaves Eq.~\eqref{eq:q_fit} invariant.
The fitted spectra determine the directional splitting magnitudes encoded by \(\bm Q_i\), while the relative microscopic Kramers frame remains free.
Signed laboratory-frame elements are represented by the unique symmetric positive-definite square root
\(\bm G_i=\bm Q_i^{1/2}\).
For an infinitesimal perturbation about a selected strain-expansion point, this canonical square root obeys the Sylvester equation
\begin{equation}
\bm G_i\delta\bm G_i
+\delta\bm G_i\bm G_i
=\delta\bm Q_i .
\label{eq:sqrt_sylvester}
\end{equation}
For a local infinitesimal response in the unperturbed principal basis,
\begin{equation}
(\delta G_i)_{ab}
=
\frac{(\delta Q_i)_{ab}}{g_{i,a}+g_{i,b}} .
\label{eq:sqrt_component_response}
\end{equation}
In that principal basis, a linear canonical \(G_{yz}\) response represents a linear physical \(Q_{yz}\) response through Eq.~\eqref{eq:sqrt_component_response}.
The complete Sylvester equation gives the corresponding relation in the laboratory/device basis.
The canonical matrices are symmetric, so \(\Delta\bm G\) has the six independent components shown in the figures.

The pair ordering is chosen continuously along each strain sweep before
\(\Delta\bm G=\bm G_1-\bm G_2\) is formed.
The first point is ordered by \(G_{1,zz}\geq G_{2,zz}\).
At each subsequent point, the same and swapped assignments have costs
\begin{align}
C_{\mathrm{same}}&=
\|\bm G_1-\bm G_1^{\mathrm{prev}}\|_{\mathrm F}^2+
\|\bm G_2-\bm G_2^{\mathrm{prev}}\|_{\mathrm F}^2,\nonumber\\
C_{\mathrm{swap}}&=
\|\bm G_2-\bm G_1^{\mathrm{prev}}\|_{\mathrm F}^2+
\|\bm G_1-\bm G_2^{\mathrm{prev}}\|_{\mathrm F}^2,
\label{eq:pair_alignment_cost}
\end{align}
and the lower-cost assignment is selected.
The labels \(1\) and \(2\) thus follow continuous tensor branches and may exchange their association with the physical fixed and tuned dots.
Signed \(\Delta\bm G\) elements and vector diagnostics use this canonical, continuity-aligned convention.
Principal values, principal axes, and local splitting magnitudes are obtained directly from the individual \(\bm Q_i\).

\bibliography{references}

\end{document}